\documentclass[iop]{emulateapj-rtx4}
\usepackage{amssymb}
\usepackage{amsmath}
\usepackage{comment}
\usepackage{hyperref}

\providecommand{\kms}{\ensuremath{\,{\rm km\,s}^{-1}}}
\providecommand{\Ang}{\ensuremath{\,\mbox{\AA}}}
\providecommand{\kpc}{\ensuremath{\,\mathrm{kpc}}}
\providecommand{\Mpc}{\ensuremath{\,\mathrm{Mpc}}}
\providecommand{\Lya}{\ensuremath{\mathrm{Ly}\alpha}} 
\providecommand{\Mgp}{\ensuremath{\mathrm{Mg\,II}}}

\providecommand{\zmgii}{\ensuremath{z_{2796}}}
\providecommand{\zqso}{\ensuremath{z_{\rm QSO}}}
\providecommand{\dv[1]}{\ensuremath{\delta v_{#1}}}
\providecommand{\dvqso}{\ensuremath{\dv{\rm QSO}}}

\providecommand{\EWr}{\ensuremath{W_{\rm r}}}

\providecommand{\EWlin[1]}{\ensuremath{\EWr{}_{,\mathrm{#1}}}}

\providecommand{\N[1]}{\ensuremath{N_{\rm #1}}}

\providecommand{\ff[1]}{\ensuremath{f(#1)}}

\providecommand{\Num}{\ensuremath{\mathcal{N}}}

\providecommand{\DX[1]}{\ensuremath{\DXp(#1)}}

\providecommand{\dNdX[1]}{\ensuremath{d \Num_{\rm #1}/d X}}
\providecommand{\dNMgIIdz}{\ensuremath{d \Num_{\Mgp}/d z}}
\providecommand{\dNMgIIdX}{\ensuremath{d \Num_{\Mgp}/d X}}
\providecommand{\dNFkIIdz}{\ensuremath{d \Num_{\mathrm{afp}}/d z}}
\providecommand{\dNFkIIdX}{\ensuremath{d \Num_{\mathrm{afp}}/d X}}

\providecommand{\sigphys}{\ensuremath{\sigma_{\rm phys}}}

\providecommand{\ncomb}{\ensuremath{n_{{\rm com,}B}}}

\begin{document}


\title{Precious Metals in SDSS Quasar Spectra II: Tracking the
  Evolution of Strong, $0.4 < z < 2.3$ \ion{Mg}{2} Absorbers with
  Thousands of Systems}

\author{Eduardo N. Seyffert\altaffilmark{1}, Kathy
  L. Cooksey\altaffilmark{2,6}, Robert A. Simcoe\altaffilmark{1}, John
  M.  O'Meara\altaffilmark{3}, Melodie M. Kao\altaffilmark{4}, and
 J. Xavier Prochaska\altaffilmark{5}}

\altaffiltext{1}{Department of Physics, MIT, 77 Massachusetts Avenue,
  37-664D, Cambridge, MA 02139, USA; enseyff@mit.edu; simcoe@space.mit.edu}

\altaffiltext{2}{MIT Kavli Institute for Astrophysics \& Space
  Research, 77 Massachusetts Avenue, 37-685, Cambridge, MA 02139, USA;
  kcooksey@space.mit.edu}

\altaffiltext{3}{Department of Chemistry and Physics, Saint Michael's
  College, One Winooski Park, Colchester, VT 05439; jomeara@smcvt.edu}

\altaffiltext{4}{Caltech, MC 249-17, 1200 East California Boulevard,
 Pasadena, CA 91125; mkao@caltech.edu}

 \altaffiltext{5}{Department of Astronomy \& UCO/Lick Observatory,
 University of California, 1156 High Street, Santa Cruz, CA 95064,
 USA; xavier@ucolick.org}

\altaffiltext{6}{NSF Astronomy \& Astrophysics Postdoctoral Fellow}

\shorttitle{Evolution of Strong \ion{Mg}{2}
  Absorbers}\shortauthors{Seyffert et al.}

\slugcomment{Draft 5: \today}

\begin{abstract}
  We have performed an analysis of over 34,000 \ion{Mg}{2} doublets at
  $0.36 < z < 2.29$ in Sloan Digital Sky Survey (SDSS) Data-Release 7
  quasar spectra; the catalog, advanced data products, and tools for
  analysis are publicly available. The catalog was divided into
  14 small redshift bins with roughly 2500 doublets in each, and from
  Monte-Carlo simulations, we estimate 50\% completeness at rest
  equivalent width $\EWr \approx 0.8\Ang$. The equivalent-width
  frequency distribution is described well by an exponential model at
  all redshifts, and the distribution becomes flatter with increasing
  redshift, i.e., there are more strong systems relative to weak
  ones. Direct comparison with previous SDSS \ion{Mg}{2} surveys
  reveal that we recover at least 70\% of the doublets in these other
  catalogs, in addition to detecting thousands of new systems. We
  discuss how these surveys come by their different results, which
  qualitatively agree but, due to the very small uncertainties, differ
  by a statistically significant amount. The estimated physical
  cross-section of \ion{Mg}{2}-absorbing galaxy halos increased
  three-fold, approximately, from $z = 0.4 \rightarrow 2.3$, while the
  $\EWr \ge 1\Ang$ absorber line density grew, \dNMgIIdX, by roughly
  45\%.  Finally, we explore the different evolution of various
  absorber populations---damped Lyman-$\alpha$ absorbers, Lyman-limit
  systems, strong \ion{C}{4} absorbers, and strong and weaker
  \ion{Mg}{2} systems---across cosmic time ($0 < z < 6$).
\end{abstract}



\section{Introduction}\label{sec.intro}


The cosmic enrichment cycle describes the movement of gas and heavy
elements (or metals) from the sites of star formation in galaxies into
the intergalactic medium (IGM) and potentially back again.  An
understanding of gas surrounding galaxies---or the circum-galactic
medium (CGM)---is crucial to understanding feedback and gas accretion
processes.  \ion{Mg}{2} $\lambda\lambda 2796, 2803$ absorption-line
surveys---in both quasar and galaxy spectra---have long been utilized
to characterize enriched, photoionized gas clouds within and
surrounding galaxies \citep[e.g.,][]{bergeron86, lanzettaetal87,
  sargentetal88, petitjeanandbergeron90, steidelandsargent92,
  churchilletal99a, weineretal09, martinandbouche09, bordoloietal11,
  lovegroveandsimcoe11, kacprzaketal11a, korneietal12,
  churchilletal13, rubinetal13}.
 
\ion{Mg}{2} is a strong transition for a wide range of ionization
parameters,\footnote{The ionization parameter, $U$, is the ratio of
  the number of H-ionizing photons to the number of H atoms.} even
with a modest number of Mg atoms, making it a common and well-studied
ion. The metallicities of \ion{Mg}{2} systems, at $0.4 < z < 1.8$,
range from one-tenth to super-solar \citep{rigbyetal02,
  charltonetal03, misawaetal08}.
Ground-based, optical spectrographs can detect the
\ion{Mg}{2} doublet over $0.1 \lesssim z \lesssim 2.5$; infrared
spectroscopy can extend that range to $z \approx 5.5$. \ion{Mg}{2}
absorption-line surveys take on three flavors: (i) galaxy
self-absorption; (ii) galaxies or quasars probing galaxies; and (iii)
quasar absorption-line (QAL) spectroscopy. Results from all
experimental setups provide evidence that \ion{Mg}{2} absorption
traces the CGM, though only the first two methods identify the host
galaxies explicitly.

Strong \ion{Mg}{2} absorbers, with rest equivalent widths of the
2796\Ang\ line $\EWlin{2796} \gtrsim 1\Ang$, have been linked to
massive, star-forming galaxies, possibly arising in their
starburst-driven outflows \citep{boucheetal06, weineretal09,
  rubinetal10}. In this model, \ion{Mg}{2} absorption is found in
cool, interlaced clumps within a heated galactic outflow. Recent
results show that the strength of the \ion{Mg}{2} absorption depends
on the azimuthal angle relative to the host galaxies
\citep{bordoloietal11, boucheetal12, kacprzaketal12}. Under the model
of a biconical starburst-driven outflow, the stronger \ion{Mg}{2}
absorbers are detected over the plane of the disk. Both studies detect
weak \ion{Mg}{2} absorption at large azimuthal angles (i.e., along the
disk axis), and these systems may trace inflowing material
\citep{chenetal10b}, possibly co-rotating with the disk
\citep{kacprzaketal11b}.

Early-type galaxies also host \ion{Mg}{2} systems \citep{chenetal10a,
  gauthieretal10, gauthieretal11}, and they tend to be weaker than
absorbers around star-forming galaxies \citep{bowenandchelouche11},
which supports the idea that weak systems may trace gas accretion or
at least the ``ambient'' CGM. Gauthier et al. estimate the covering
fraction of \ion{Mg}{2} absorption to be $\approx\!14\%$ for massive
luminous red galaxies and greater for less massive
galaxies. \citet{chenetal10a} searched for absorbers in quasar
sightlines near galaxies (as opposed to seeking galaxies near
sightlines with known \ion{Mg}{2} systems) and found no correlation to
the star-formation rate of the host galaxies. They traced the
absorption in bright, field galaxies out to about 100\kpc\ with
50--80\% covering fraction.

\citet{gauthier13} propose that ultra-strong, $\EWlin{2796} \gtrsim
3\Ang$ \ion{Mg}{2} systems trace gas in galaxy groups \citep[though
see][]{nestoretal11}. The ultra-strong \ion{Mg}{2} absorbers have
broad, kinematically complex profiles. Galaxies in groups may have
different radial \EWlin{2796}\ profiles from isolated galaxies
\citep{chenetal10a}.

\ion{Mg}{2} absorption has been used to select damped \Lya\ absorber
(DLA) candidates at $z \lesssim 1.6$, because it is a common
transition for even moderately metal-enriched gas and can be observed
with ground-based optical spectrographs \citep{raoandturnshek00},
whereas the distinctive \Lya\ profile is accessible only with
space-based ultraviolet spectrographs. DLAs, with \ion{H}{1} column densities
$\log N_{\rm H\,I} \ge 20.3$, are considered the cold gas reservoirs
for star formation and thought to reside in star-forming galaxies
\citep[see][and references therein]{wolfeetal05}.

The large database of quasar spectra generated by SDSS
\citep{yorketal00} provides a way to efficiently find strong
\ion{Mg}{2} systems in the range $0.4 < z < 2.3$, or from when the
universe was 9.5\,Gyr to 3\,Gyr old. At SDSS resolution
($\approx\!150\kms$) and typical signal-to-noise, SDSS is more
complete at $\EWlin{2796} \gtrsim 1\Ang$, defining ``strong''
\ion{Mg}{2} absorbers. There have been four SDSS \ion{Mg}{2} surveys in:
the early data release or
EDR---\hspace{-2pt}\defcitealias{nestoretal05}{N05}\citet[][hereafter
N05]{nestoretal05};
DR3---\hspace{-2pt}\defcitealias{prochteretal06}{P06}\citet[][P06]{prochteretal06};
DR4---\hspace{-2pt}\defcitealias{quideretal11}{Q11}\citet[][Q11]{quideretal11};
and
DR7---\hspace{-2pt}\defcitealias{zhuandmenard13}{ZM13}\citet[][ZM13]{zhuandmenard13}.\footnote{\citet{lundgrenetal09}
  conducted a completely automated \ion{Mg}{2} survey in a sub-sample
  of DR5 quasar spectra, in order to conduct a \ion{Mg}{2}-galaxy
  clustering analysis. We exclude comparison to this targeted QAL
  survey; however, \citetalias{zhuandmenard13} compare with the
  \citet{lundgrenetal09} results.\label{fn.lundgren}}

Most of these studies measured the frequency distribution of
\EWlin{2796} and the absorber redshift density, \dNMgIIdz. The former
is fit well by an exponential model, showing that there is a break in
the full equivalent-width distribution, when accounting for the weaker
systems being well-modeled by a power-law distribution
\citep{churchilletal99a, narayananetal07}. For $\EWlin{2796} \ge
1\Ang$, \dNMgIIdz\ increases by a factor of $\approx\!2.5$ from $z
\approx 0.4 \rightarrow 2.3$. Recent results from an infrared survey
showed that \dNMgIIdz\ decreases from $z \approx 2 \rightarrow 5.5$
\citep{matejekandsimcoe12}.

\citetalias{prochteretal06} first connected the \ion{Mg}{2} evolution
to the cosmic star-formation rate density, $\rho_{\ast}$, and
\citet{menardetal11} constructed an empirical relation between
\dNMgIIdz\ and $\rho_{\ast}$, which peaks at $z \approx 2$ to 3 and
roughly matches the \dNMgIIdz\ evolution
(\citealt{matejekandsimcoe12}, \citetalias{zhuandmenard13}).  Though
QAL surveys do not produce direct ties between \ion{Mg}{2} absorbers
and galaxies, the seemingly related evolution of \dNMgIIdz\ for
$\EWlin{2796} \ge 1\Ang$ and $\rho_{\ast}$ is at least circumstantial
evidence that strong \ion{Mg}{2} systems are linked to star
formation. Also, \citet{menardetal11} detected [\ion{O}{2}]
$\lambda3727$ nebular emission in the SDSS fibers with \ion{Mg}{2}
absorption, which directly ties the absorption to star formation.

\citet{matejekandsimcoe12} observed little evolution in \dNMgIIdz\ for
$0.3\Ang \le \EWlin{2796} < 1\Ang$ systems over $0.4 \lesssim \zmgii
\lesssim 5.5$, comparing to \citetalias{nestoretal05} and later upheld
by \citetalias{zhuandmenard13}. This suggests a population of weak
\ion{Mg}{2} absorbers being established early in the history of the
universe or being constantly replenished, at a conserving rate, over
time.

In spite of the numerous observations of \ion{Mg}{2} absorbers and
their direct or inferred relationship with galaxies, the origin of the
absorbing gas in the CGM is not yet clearly understood.
We are motivated to conduct a SDSS \ion{Mg}{2} survey so that we may
fairly compare the results with our high-redshift results
\citep{matejekandsimcoe12} and with our other SDSS metal-line
surveys. We have completed the search for \ion{C}{4}
$\lambda\lambda1548,1550$ \defcitealias{cookseyetal13}{Paper
  I}\citep[][hereafter, Paper I]{cookseyetal13}, and our survey of
\ion{Si}{4} $\lambda\lambda1393,1402$ is in preparation. We also
discuss why the various SDSS \ion{Mg}{2} surveys produce 
different results, which, due to the large sample sizes, are
statistically significant.

We summarize how we construct our \ion{Mg}{2} catalog, correct for
completeness, and compare with previous SDSS catalogs in Section
\ref{sec.catalog}, while absorber-by-absorber comparisons are left to
Appendix \ref{appdx.prevcat}. The main results are detailed in Section
\ref{sec.results}. We discuss the implications of our results in the
context of the CGM and the results from surveys of other absorber
populations in Section \ref{sec.discuss}. The summary is given in
Section \ref{sec.summ}. We adopt the WMAP5 cosmology:
$H_{0}=71.9\kms\Mpc^{-1}$, $\Omega_{\rm M} = 0.258$, and
$\Omega_{\Lambda} = 0.742$ \citep{komatsuetal09}.

\begin{deluxetable*}{llllrccccc}
\tablewidth{0pc}
\tablecaption{Sightline Summary\label{tab.los}}
\tabletypesize{\scriptsize}
\tablehead{ 
\colhead{(1)} & \colhead{(2)} & \colhead{(3)} & 
\colhead{(4)} & \colhead{(5)} & 
\colhead{(6)} & \colhead{(7)} & \colhead{(8)} & 
\colhead{(9)} & \colhead{(10)} \\ 
\colhead{QSO ID} & \colhead{R.A.} & 
\colhead{Decl.} & \colhead{\zqso} & 
\colhead{$\langle {\rm S/N} \rangle$} & \colhead{$f_{\rm BAL}$} & 
\colhead{$\Delta X_{\rm max}$} & 
\colhead{$\Num_{\rm cand}$} & \colhead{$\Num_{\rm Mg\,II}$} & 
\colhead{$\delta X_{\rm Mg\,II}$} \\ 
 & & & & 
\colhead{(pixel$^{-1}$)} &  &  &  & & 
} 
\startdata 
52203-0685-467 & 00:00:06.53 &  +00:30:55.2 &  1.8246 &  7.85 &   0 &   4.04 &   0 &   0 & \nodata \\ 
52203-0685-470 & 00:00:08.13 &  +00:16:34.6 &  1.8373 & 10.34 &   0 &   2.27 &   0 &   0 & \nodata \\ 
52235-0750-082 & 00:00:09.38 &  +13:56:18.4 &  2.2342 &  4.17 &   0 &   0.19 &   2 &   0 & \nodata \\ 
52143-0650-199 & 00:00:09.42 & --10:27:51.9 &  1.8449 &  8.54 &   0 &   3.33 &   2 &   1 & 0.028 \\ 
52235-0750-499 & 00:00:11.41 &  +14:55:45.6 &  0.4597 &  7.83 &   0 &   0.49 &   0 &   0 & \nodata \\ 
51791-0387-200 & 00:00:11.96 &  +00:02:25.3 &  0.4789 & 11.16 &   0 &   1.64 &   0 &   0 & \nodata \\ 
52235-0750-098 & 00:00:13.14 &  +14:10:34.6 &  0.9582 & 12.26 &   0 &   2.15 &   0 &   0 & \nodata \\ 
52203-0685-198 & 00:00:14.82 & --01:10:30.7 &  1.8877 &  9.21 &   0 &   2.87 &   0 &   0 & \nodata \\ 
52203-0685-439 & 00:00:15.47 &  +00:52:46.8 &  1.8516 &  7.22 &   0 &   2.47 &   2 &   0 & \nodata \\ 
52991-1489-142 & 00:00:16.43 & --00:18:33.3 &  0.7030 &  6.20 &   0 &   0.92 &   0 &   0 & \nodata \\ 
52143-0650-459 & 00:00:17.38 & --08:51:23.7 &  1.2491 &  7.44 &   0 &   3.74 &   1 &   0 & \nodata
\enddata 
\tablecomments{ 
Column 1 is the adopted QSO identifier from the spectroscopic modified Julian date, plate, and fiber number.
Columns 2 through 4 are from the DR7 QSO catalog \citep{schneideretal10}.
Column 5 is the median S/N measured in the region searched for Mg\,II.
The binary BAL flag $f_{\rm BAL}$ in Column 6 indicates which sightlines were considered BALs by at least one author (4) and which were confirmed by the authors as BALs to exclude (8).
Column 7 is the maximum co-moving pathlength available in the sightline.
Columns 8 and 9 give the number of candidate and confirmed Mg\,II doublets, respectively.
Column 10 is the pathlength blocked by the $\Num_{\rm Mg\,II}$ doublets in the sightline.
(This table is available in a machine-readable form in the online journal.
A portion is shown here for guidance regarding its form and content.)
} 
\end{deluxetable*} 

\begin{figure}[hbt]
  \begin{center}
    \includegraphics[width=0.49\textwidth]{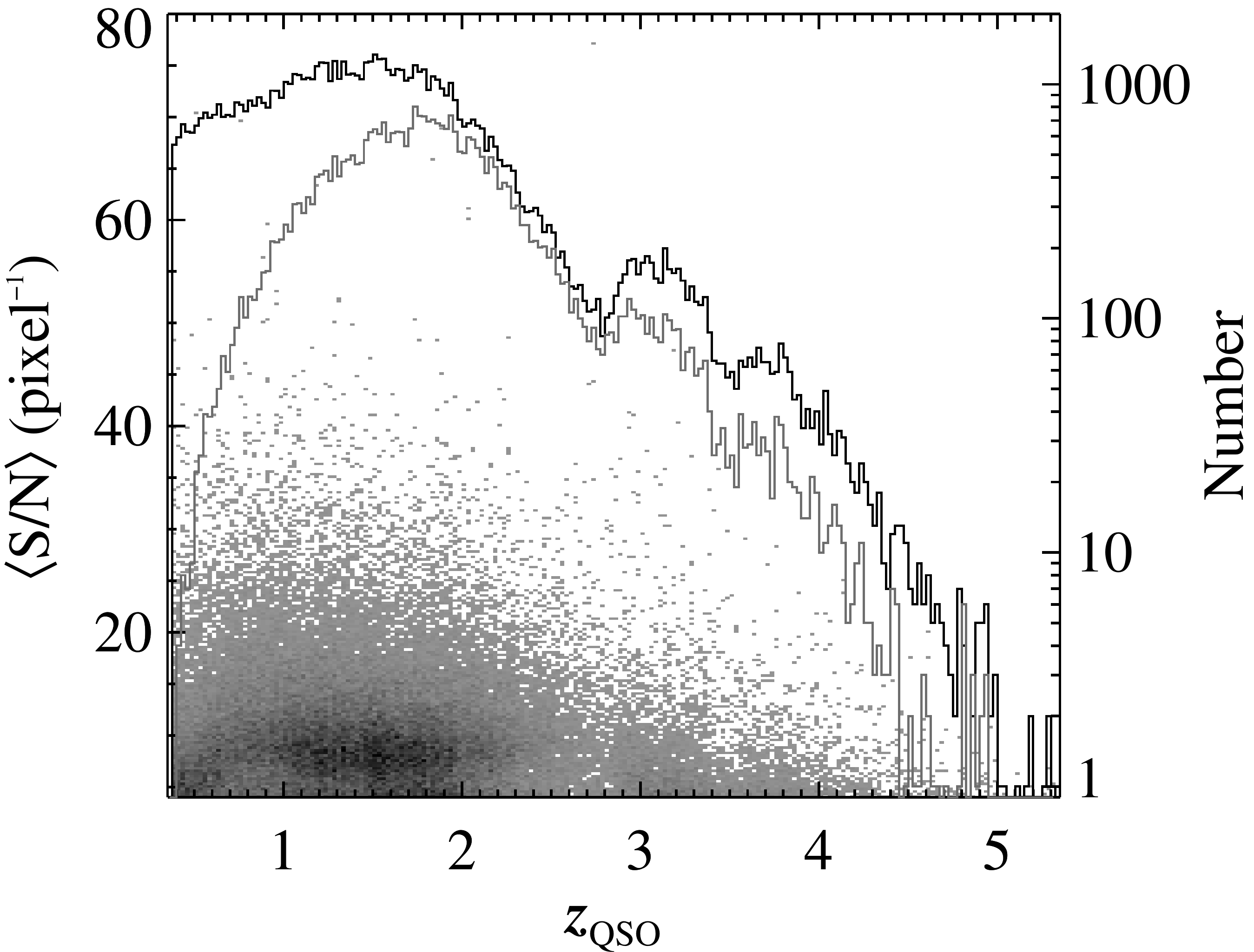}
  \end{center}
  \caption[Redshift and $\langle {\rm S/N} \rangle$ distribution of
  sightlines]
  {Redshift and $\langle {\rm S/N} \rangle$ distribution of
    sightlines. Using the left-hand axis, the 2D histogram shows the
    median signal-to-noise and \zqso\ space of the analyzed 79,294
    spectra. The black and gray histograms give the redshift
    distribution for all spectra and for the 28,938 with confirmed
    \ion{Mg}{2} absorbers, respectively (right-hand axis).
    \label{fig.zqsosnr}
  }
\end{figure}


\section{Constructing the \ion{Mg}{2} Catalog}\label{sec.catalog}

\subsection{From Quasar Spectra to Visually Verified
  Sample}\label{subsec.methodsumm}

Our \ion{Mg}{2} absorber catalog was constructed using a subset of the
SDSS DR7 quasar catalog \citep{schneideretal10} and following an
equivalent methodology to the \ion{C}{4} survey described in detail in
\citetalias{cookseyetal13}, to which the interested reader is
referred. Here we briefly outline the procedure.

Of the 105,783 DR7 quasar spectra, 79,595 were searched for
\ion{Mg}{2} systems (Table \ref{tab.los}) because they: (i) were not
broad absorption-line (BAL) QSOs \citep[identified by][]{shenetal11}
and (ii) had median signal-to-noise ratio $\langle {\rm S/N} \rangle
\ge 4\,{\rm pixel}^{-1}$ in the region covering intergalactic
\ion{Mg}{2} absorption (i.e., observed wavelengths greater than
$1250\Ang(1+\zqso)$ and $\dvqso < -3000\kms$).\footnote{Velocity
  offset is defined as $\dvqso = c(\zmgii - \zqso)/(1 +\zqso)$.} A
further 301 sightlines were later excluded as ``visual'' BAL QSOs,
leaving a total of 79,294 quasar spectra included in this survey;
their distribution in \zqso--$\langle {\rm S/N} \rangle$ space is shown
in Figure \ref{fig.zqsosnr}.

Every quasar spectrum was normalized with its ``hybrid continuum,'' a
fit combining principle-component analysis, b-spline correction, and
pixel\slash absorption-line rejection. Absorption-line candidates were
automatically detected by convolving the normalized flux and error
arrays with a Gaussian kernel with a full-width at half-maximum of one
pixel, roughly an SDSS resolution element (resel). Candidate
\ion{Mg}{2} doublets were identified based solely on the
characteristic velocity separation ($767\kms$), plus/minus $150\kms$
to allow for blending. The candidate doublets had convolved $({\rm
  S/N})_{\rm conv} \ge 3.5\,{\rm resel}^{-1}$ in the 2796\Ang\ line
and $2.5\,{\rm resel}^{-1}$ in the 2803\Ang\ line. Any automatically
detected absorption feature with $({\rm S/N})_{\rm conv} \ge 3.5\,{\rm
  resel}^{-1}$ and broad enough to enclose a \ion{Mg}{2} doublet was
included in the candidate list.

The wavelength bounds of the absorption lines were automatically
defined by where the convolved S/N array began
increasing when stepping away from the automatically detected line
centroid. The new centroid was then set as the flux-weighted average
wavelength within the bounds, and the \ion{Mg}{2} doublet redshift,
\zmgii, was defined by the new 2796\Ang\ centroid.

From simulations of synthetic \ion{Mg}{2} profiles with known
equivalent widths, we empirically determined that capping the flux at
the continuum plus one sigma (Poisson uncertainty of the flux and
continuum error) yielded the most accurate \EWr\ measurements using
the technique of boxcar summation; the measured \EWr\ was 0.08\Ang\
larger than the input \EWr (in the median), with a median absolute
deviation of 0.21\Ang.  Hence, our (observed) equivalent widths are
the sum of this modified absorption within the wavelength bounds
(i.e., $(1-f)\delta\lambda$). 

All candidates were visually inspected by at least one author and most
by two. They were rated on a four-point scale from 0 (definitely
false) to 3 (definitely true). The systems were judged largely on the
\ion{Mg}{2} doublet (e.g., centroid alignment, correlated profiles)
but possibly associated ions were also reviewed to aid
verification. All \ion{Mg}{2} absorbers with rating of 2 or 3 were included in
subsequent analyses and combined into systems if separated by less
than $250\kms$.

Ultimately, we detected 35,629 doublets---from over 90,000
candidates---with $\EWlin{2796} < 9\Ang$ and $0.36 < \zmgii < 2.29$
(Table \ref{tab.mgii}). 
We excluded the \ion{C}{4} emission-line region, $\pm3000\kms$ around
$1548.195\Ang(1+\zqso)$, because the continuum fits in this region
were questionable, owing to the potentially large velocity offsets of
the emission lines and the large incidence of strong intrinsic
absorption. In addition, real intervening \ion{Mg}{2} absorption could
be lost in the intrinsic \ion{C}{4} absorption, thus hindering the
recovery of \ion{Mg}{2} systems in this region. This reduced the total
sample size by 531. We also limit our sample to systems with $\dvqso <
-5000\kms$, leaving 34,254 doublets for further analyses.

\begin{deluxetable*}{rllr@{\,$\pm$\,}lr@{\,$\pm$\,}lr@{\,$\pm$\,}l}
\tablewidth{0pc}
\tablecaption{\ion{Mg}{2} System Summary\label{tab.mgii}}
\tabletypesize{\scriptsize}
\tablehead{ 
\colhead{(1)} & \colhead{(2)} & \colhead{(3)} & 
\multicolumn{2}{c}{(4)} & \multicolumn{2}{c}{(5)} & 
\multicolumn{2}{c}{(6)}  \\ 
\colhead{QSO ID} & \colhead{\zqso} & 
\colhead{$z_{2796}$} & 
\multicolumn{2}{c}{\EWlin{2796}} & \multicolumn{2}{c}{\EWlin{2803}} & 
\multicolumn{2}{c}{$C(\EWlin{2796})$} \\ 
 & & & 
\multicolumn{2}{c}{(\AA)} & \multicolumn{2}{c}{(\AA)} & \multicolumn{2}{c}{} } 
\startdata 
52235-0750-082 &  2.2342 & 0.96363 &  2.445 &  0.174 &  1.547 &  0.168 &  0.915 &  0.006 \\ 
                         &           & 1.14500 &  0.268 &  0.115 &  0.254 &  0.143 &  0.112 &  0.069 \\ 
52143-0650-199 &  1.8449 & 1.31252 &  1.965 &  0.235 &  2.016 &  0.198 &  0.882 &  0.034 \\ 
                         &           & 1.52824 &  0.394 &  0.159 &  0.507 &  0.180 &  0.203 &  0.113 \\ 
52203-0685-439 &  1.8516 & 1.37148 &  0.650 &  0.212 &  1.140 &  0.253 &  0.400 &  0.163 \\ 
                         &           & 1.63660 &  1.457 &  0.577 &  2.376 &  0.396 &  0.789 &  0.249 \\ 
52143-0650-459 &  1.2491 & 0.69284 &  0.575 &  0.162 &  0.486 &  0.164 &  0.344 &  0.126 \\ 
54389-2822-318 &  1.1560 & 0.48941 &  2.724 &  0.384 &  2.534 &  0.375 &  0.924 &  0.012 \\ 
51791-0387-167 &  2.1249 & 1.07105 &  0.801 &  0.256 &  0.631 &  0.240 &  0.495 &  0.174 \\ 
51791-0387-531 &  0.9511 & 0.49645 &  0.310 &  0.115 &  0.678 &  0.132 &  0.142 &  0.074 \\ 
54389-2822-339 &  1.4446 & 0.38268 &  1.006 &  0.555 &  1.684 &  0.564 &  0.600 &  0.352 \\ 
                         &           & 0.76378 &  2.117 &  0.309 &  1.581 &  0.304 &  0.898 &  0.038 \\ 
                         &           & 1.09345 &  0.852 &  0.387 &  1.036 &  0.406 &  0.526 &  0.268 \\ 
52143-0650-178 &  2.6404 & 1.05622 &  1.749 &  0.141 &  1.283 &  0.172 &  0.851 &  0.025 \\ 
                         &           & 1.38095 &  0.793 &  0.113 &  0.619 &  0.116 &  0.490 &  0.071 \\ 
52251-0751-355 &  1.4115 & 0.89761 &  1.749 &  0.297 &  1.799 &  0.290 &  0.852 &  0.064 \\ 
51791-0387-093 &  1.8973 & 1.02077 &  0.723 &  0.172 &  0.791 &  0.214 &  0.446 &  0.120
\enddata 
\tablecomments{ 
For each sightline (identified in Columns 1 and 2), every confirmed doublet is listed by the redshift of its Mg\,II 2796 line (Column 3).
The rest equivalent widths of the Mg\,II lines are given in Columns 4 and 5.
In Column 6, we give the completeness fraction for the doublet from the whole survey average.
(This table is available in a machine-readable form in the online journal.
A portion is shown here for guidance regarding its form and content.)
}
\end{deluxetable*} 

\begin{figure*}[hbt]
  \begin{center}$
    \begin{array}{cc}
  \includegraphics[width=0.47\textwidth]{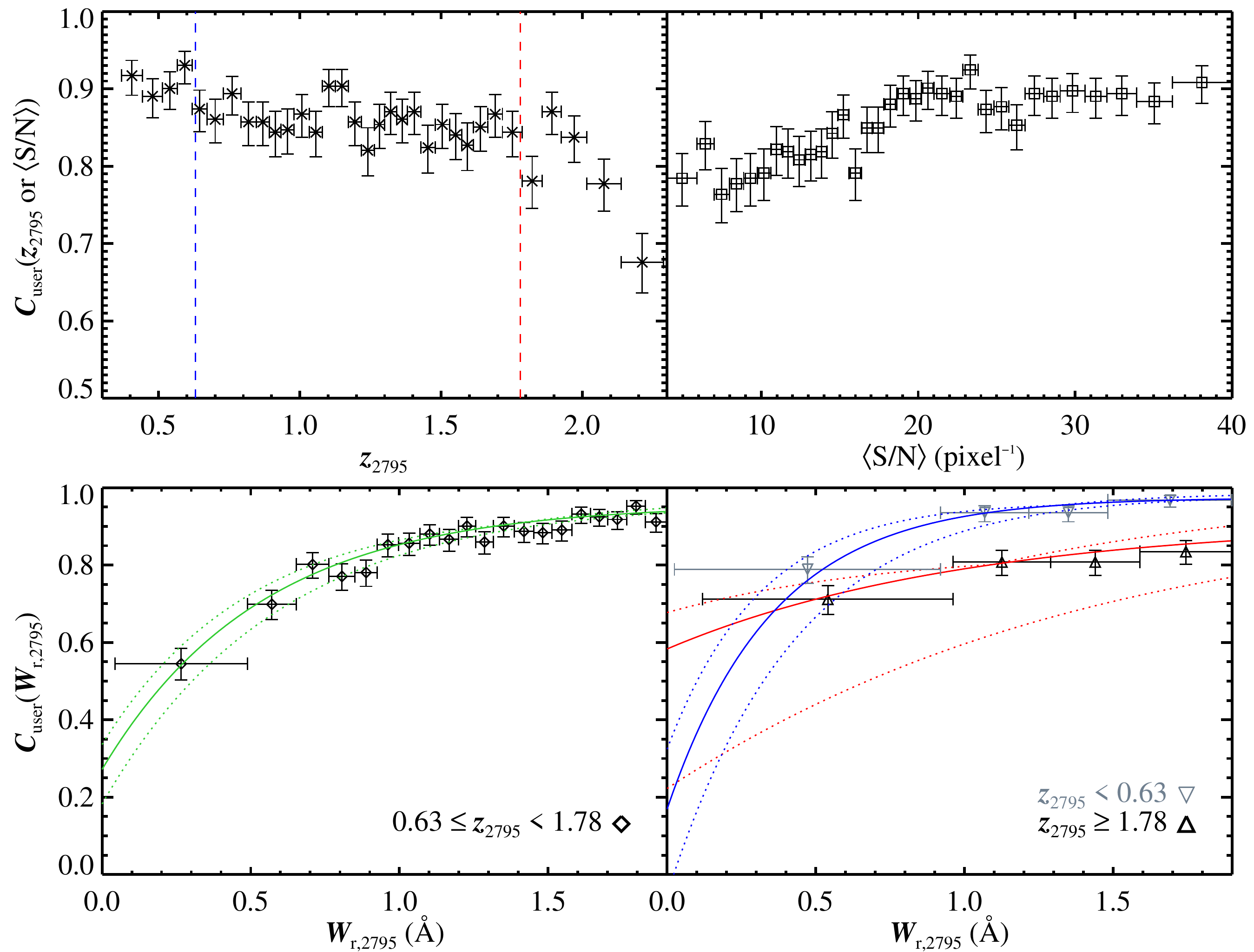}  & 
  \includegraphics[width=0.47\textwidth]{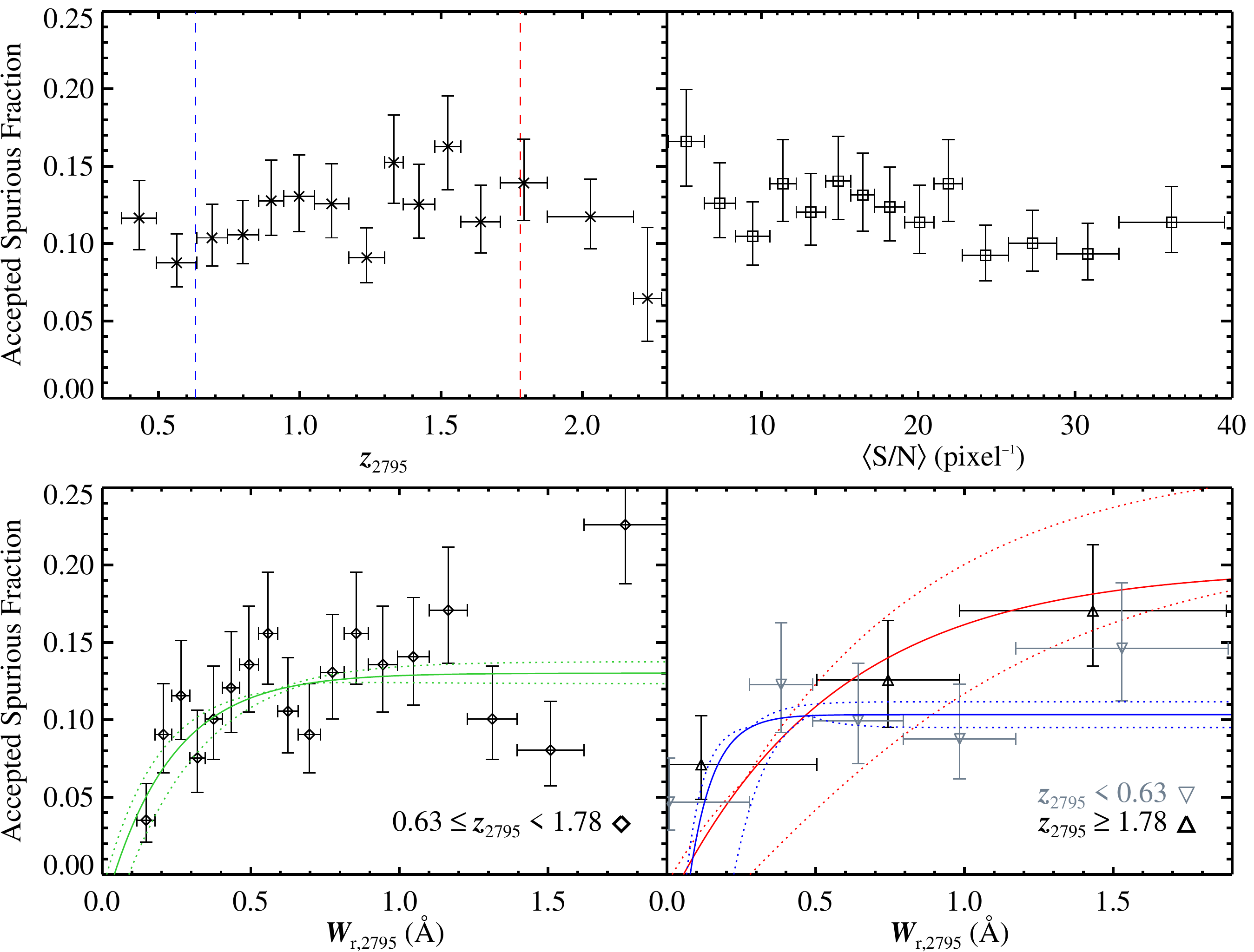} 
  \end{array}
 $\end{center}
  \caption[Biases of visual verification and trends of spurious detections.]
  {Biases of visual verification and trends of spurious
    detections. {\it Left}: We tested the ``user bias'' by rating fake
    $\lambda\lambda2795,2801$ doublets that were injected and
    automatically recovered as candidates. The four panels show the
    completeness of our ability to correctly rate true doublets, as a
    function of redshift, spectrum $\langle {\rm S/N} \rangle$, and
    equivalent width in three redshift bins. The redshift cuts (vertical
    lines) were determined by changes in $C_{\rm user}(z_{2795})$
    and to align with the closest redshift bins used in our
    analyses. The solid and dashed lines are the best-fit model,
    $C_{\rm user}(\EWr) = C_{0}(1-\exp(\beta(\EWr-W_{0})))$, and its
    1-$\sigma$ errors. {\it Right}: We plot the accepted
    false-positive fraction as functions of redshift, signal-to-noise,
    and equivalent width. These spurious pairs of lines were {\it not}
    injected but were automatically recovered. The lower panels show
    the best-fit model and errors to the accepted spurious fraction
    over \EWlin{2795}.
    \label{fig.userbias}
  } 
\end{figure*}

\subsection{Testing and Correcting for
  Completeness}\label{subsec.cmplt}

As in \citetalias{cookseyetal13}, we test our survey completeness with
Monte-Carlo simulations. Ultimately, our completeness fractions,
$C(z,W)$, are the ratios of the number of {\it accepted} (i.e.,
visually verified), simulated doublets, $\Num_{\rm accept}$, to the
number input, $\Num_{\rm input}$, as a function of \zmgii\ and
\EWlin{2796}. We estimate $C(z,W)$ in four steps, described below.

First, in a ``basic'' completeness test, we measured the number,
$\Num_{\rm rec}$, of simulated \ion{Mg}{2} doublets recovered by our
automated procedures, from continuum fitting to candidate
selection. We generated a library of synthetic profiles, inserted them
into a representative subset of sightlines, and tracked which profiles
were automatically recovered. The simulated doublets reflected the
observed variety in our SDSS \ion{Mg}{2} catalog. By construction,
they uniformly spanned $0.05\Ang \le \EWlin{2796} < 6.5\Ang$ but
$\approx\!20\%$ sampled 6.5\Ang--8\Ang\ with decreasing frequency. We
injected 1000 simulated doublets in 30\% of the sightlines at a rate
of $d\Num/dz = 5$. The tested sightlines sampled the entire
\zqso--$\langle {\rm S/N} \rangle$ space, and the un-sampled
sightlines were assigned the mean completeness fraction of the tested
sample in small bins of \zqso\ and $\langle {\rm S/N} \rangle$.

Second, we estimated the ``user'' bias, resulting from our visual
verification. We determined the rate at which we accepted real
doublets (``true positives'') and spurious pairs (``accepted false
positives''); the latter are chance alignments and\slash or noise
fluctuations. As in \citetalias{cookseyetal13}, we injected simulated
fake doublets, with rest wavelength comparable to \ion{Mg}{2} but with
characteristic separation of 652\kms, and visually rated the recovered
candidates, thus measuring the accepted number, $\Num_{\rm
  accept}$. Effectively, the completeness fraction is now:
\begin{equation}
  C(z,W) = \frac{ \Num_{\rm rec}(z,W)
  } { \Num_{\rm input}(z,W) }\frac{ \Num_{\rm accept}(z,W)
  } { \Num_{\rm rec}(z,W) } {\rm .} \label{eqn.czw}
\end{equation}
However, as in \citetalias{cookseyetal13}, we fitted a functional form to
the accepted-to-recovered fraction, $C_{\rm user}(z,W) =
C_{0}(1-\exp(\beta(\EWr-W_{0})))$. This adjustment slightly increased
the total completeness fraction measured in the ``basic'' test.

We injected $\Num_{\rm input} = 16,209$ fake doublets with $\EWr <
2\Ang$ and $\dvqso < -5000\kms$, and $\Num_{\rm rec} = 10,258$ were
automatically recovered. Of these, we correctly rated $\Num_{\rm
  accept} = 8472$. The automatically generated candidate list
included any pair of lines with separations within $\pm150\kms$ (see
Section \ref{subsec.methodsumm}) of the fake doublet separation, which
brought forth true fake doublets, real \ion{Mg}{2} systems, and common
contaminants. These contaminating pairs of lines intrinsically had a
range of separations---fixed, if they are from the same system, or
random, if a pair of spurious lines. Thus the $652\pm150\kms$ search
window was sampled by a distribution very representative of
contaminants affecting the \ion{Mg}{2} survey.
 
As shown in Figure \ref{fig.userbias} (left panels), the user
completeness generally decreases with increasing redshift and
decreasing spectrum $\langle {\rm S/N} \rangle$ and \EWr. We divide
the $C_{\rm user}$ fit at $z = 0.63$ and $z = 1.78$, where $C_{\rm
  user}$ changes rapidly and which align with our fixed bins.

\begin{figure*}[hbt]
  \begin{center}
    \includegraphics[width=0.94\textwidth]{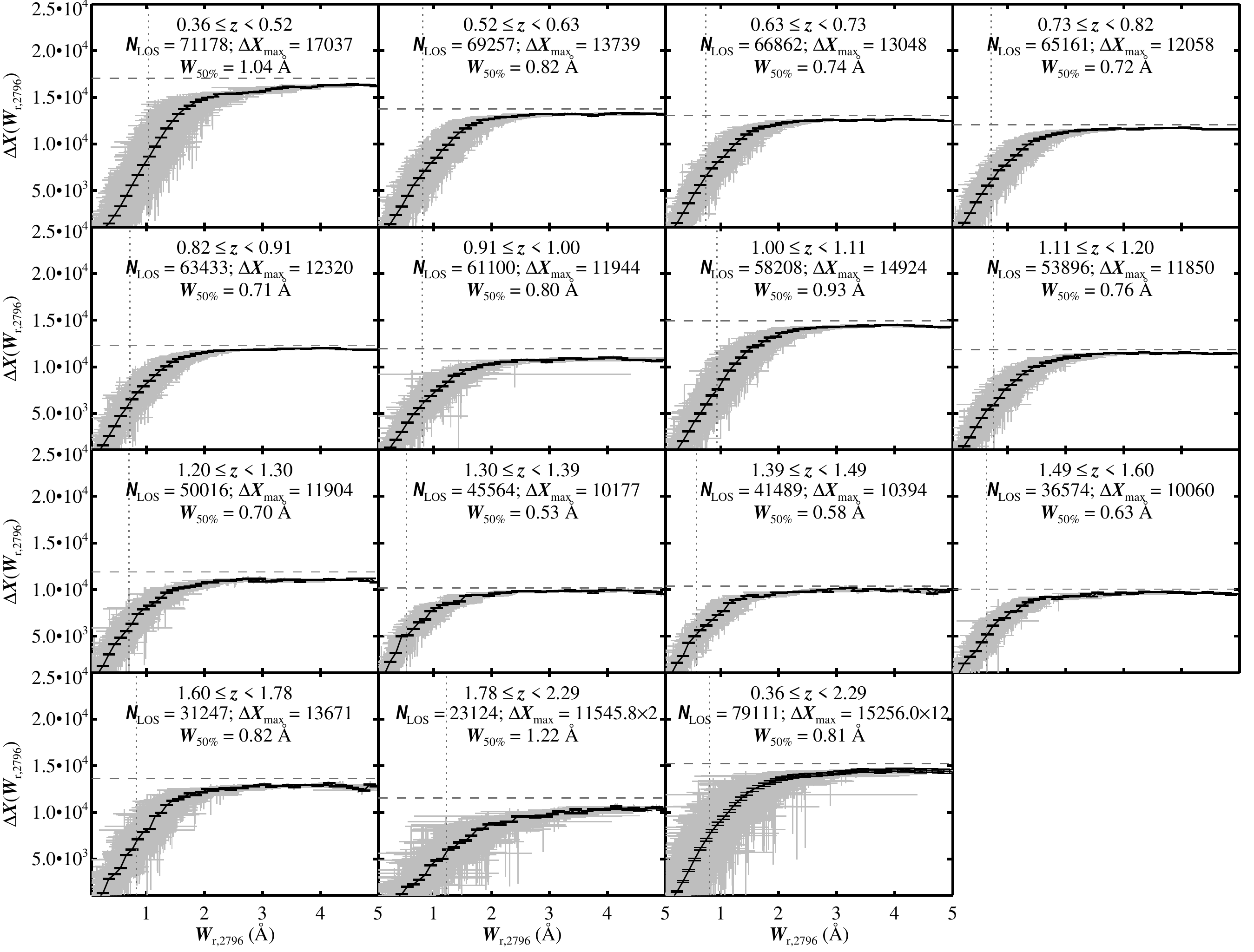}
  \end{center}
  \caption[Results from Monte-Carlo completeness tests.]
  {Results from Monte-Carlo completeness tests. The
    completeness-corrected co-moving path length as a function of rest
    equivalent width, \DX{\EWlin{2796}}, is shown for the 15 redshift
    bins used in the current study. The black lines and errors are the
    completeness curves, and the gray points and errors are the
    observations. The horizontal, dashed line traces the maximum path
    length available in the bin, and the vertical, dotted line
    indicates the equivalent width where we are 50\% complete.
    \label{fig.cmplt}
  }
\end{figure*}

Third, we scaled the completeness by the fraction of the total
co-moving path length\footnote{The co-moving path length is related to
  redshift as follows: $dX/dz = (1+z)^2/\sqrt{ \Omega_{\rm M}(1+z)^{3}
    + \Omega_{\Lambda}}$.\label{fn.dxdz}} {\it not} blocked by
doublets with greater \EWr\ in the given redshift bin. All \ion{Mg}{2}
lines blocked less than 1\% of the total survey path length (Table
\ref{tab.mgii}). This correction slightly decreased the total
completeness fraction. To measure the unblocked co-moving path length
available in our survey, we multiply $C(W)$ by the total path
available in the given redshift bin (Figure \ref{fig.cmplt}).

Fourth, the user completeness tests enabled measurement of the
accepted false-positive rate, \dNFkIIdX.  There were 5802 spurious
candidates from the automated search, of which we incorrectly accepted
678. As shown in Figure \ref{fig.userbias} (right panels), the
accepted false-positive fraction is roughly constant at $\approx10\%$
as a function of redshift and signal-to-noise. However, the fraction
increases with increasing \EWr, plateauing to $\approx\!10\%$ at $z <
1.78$ and $\approx\!20\%$ at higher redshift.  For $\EWr \ge 1\Ang$,
we estimate $\dNFkIIdX = 0.0015$ ($z < 0.63$), 0.0020 ($0.63 \le z <
1.78$), and 0.0045 ($z \ge 1.78$). We explain how we adjust for the
accepted false-positive rate in Section \ref{sec.results}.

As mentioned previously, we excluded the $\pm3000\kms$ around the
quasar \ion{C}{4} emission line. This reduced the pathlength by 7\%
over $1.48 \le z < 1.78$ and less (2\%--5\%) in other noticeably
affected bins, $1.20 \le z < 2.29$.

\begin{figure*}[thb]
  \begin{center}
    \includegraphics[width=0.94\textwidth]{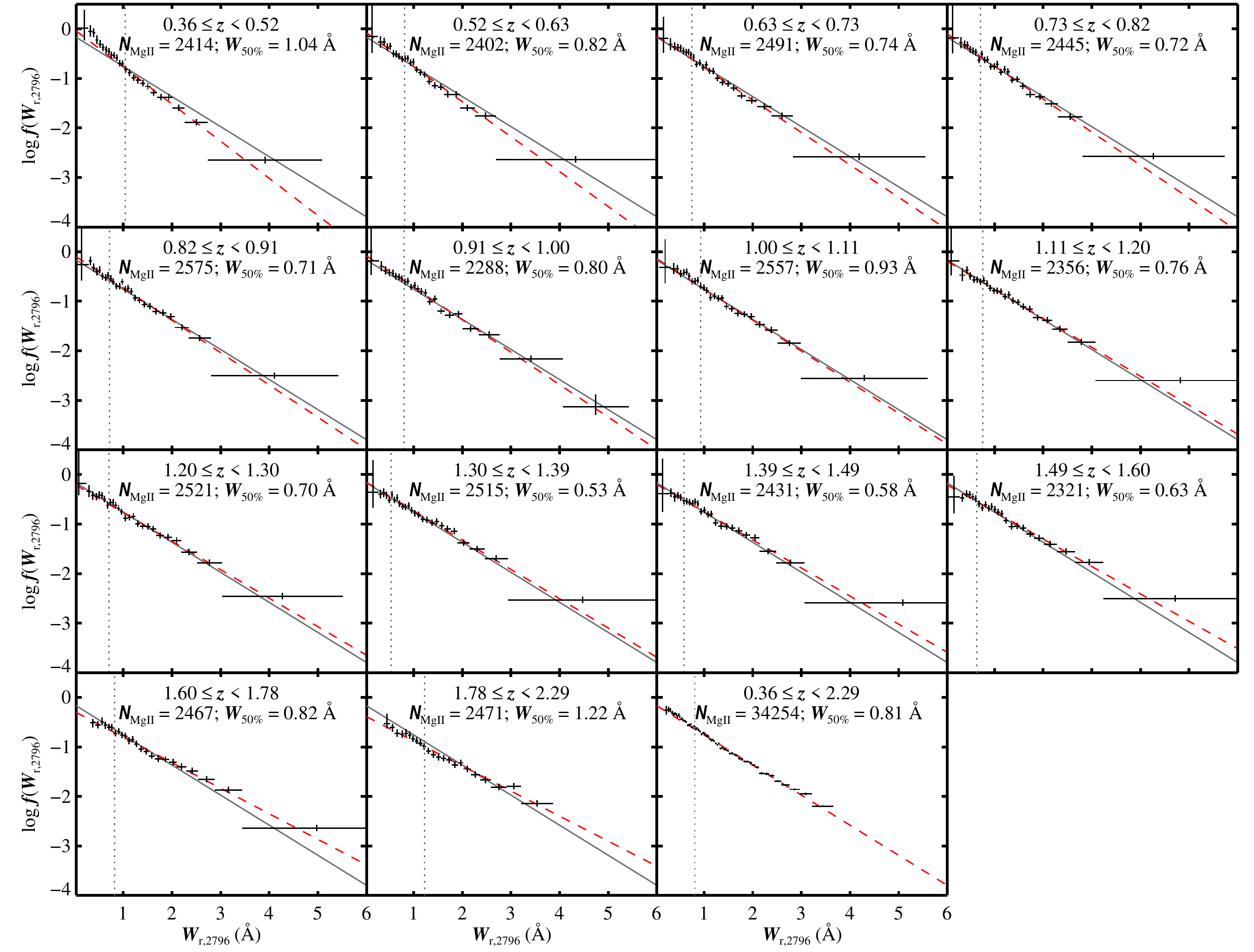}
  \end{center}
  \caption[Equivalent-width frequency distributions.]
  {Equivalent-width frequency distributions. The maximum likelihood
    fits of an exponential function are the dashed (red) lines, for
    each redshift bin, and the solid (gray) line, for the full
    sample. There is distinct evolution with redshift: \ff{\EWr}\
    flattens (larger fraction of strong absorbers) and increases (more
    absorbers overall) with increasing redshift. The observations have
    been completeness corrected, and the redshift-specific 50\%
    completeness limits are the vertical dotted lines.
    \label{fig.fw}
  }
\end{figure*}

\subsection{Comparing with Previous SDSS \ion{Mg}{2}
  Catalogs}\label{subsec.prevcat}

There have been four \ion{Mg}{2} surveys using different SDSS data
releases: early \citep{nestoretal05}; third \citep{prochteretal06};
fourth \citep{quideretal11}; and seventh
\citep{zhuandmenard13}.\footnote{See footnote \ref{fn.lundgren}.} The surveys
have a variety of differences that contribute to variations between
line lists for the same subset of quasars and, hence, the results
(e.g., \dNMgIIdX), which we discuss in Section \ref{sec.results}. Here
we summarize the differences in catalogs and methodologies---from
continuum fitting to final doublet selection; we leave the detailed
absorber-to-absorber comparisons to Appendix \ref{appdx.prevcat}.


\citetalias{nestoretal05} and \citetalias{quideretal11} model the
intrinsic quasar spectra with a combination of cubic splines and
Gaussian profiles for emission lines, while
\citetalias{prochteretal06} use b-splines and principle-component
analysis (PCA) for emission lines. \citetalias{zhuandmenard13} and
this study use PCA for both continua and emission lines and correct
for low-frequency modulations automatically in post-processing.

All catalogs are visually verified, except for
\citetalias{zhuandmenard13}. The latter uses the
\citetalias{quideretal11} catalog as their training set, and
\citetalias{quideretal11}, in turn, is based on the methodologies of
\citetalias{nestoretal05}. These three studies model the absorption
lines with a (single or double) Gaussian profile, from which they
measure the equivalent widths. These three studies and our work agree
well on \EWlin{2796} values.

Our automated candidate doublet search is based on the algorithm used
by \citetalias{prochteretal06}.  We both use boxcar summation to
measure equivalent widths, but \citetalias{prochteretal06} fix the box
width, while we let it vary automatically depending on the
profile. The fixed width contributes to the disagreement in our
equivalent-width measurements, with \citetalias{prochteretal06}
underestimating \EWlin{2796}. More importantly, the continuum
placement in \citetalias{prochteretal06} was biased low by the strong
absorption systems, systematically reducing their equivalent widths.

All surveys estimate their sample completeness, and all but
\citetalias{quideretal11} use Monte-Carlo simulations in a similar
fashion to ours, though number of realizations, types of profiles,
etc.  differ. \citetalias{quideretal11} relied on the tests of the
automated algorithms that \citetalias{nestoretal05} conducted and
assessed their false-negative rate by having two authors visually
verify a small number of sightlines. \citetalias{quideretal11} focused
on describing their catalog, and any analyses did not rely on
correcting for completeness.

To summarize, the detailed catalog-to-catalog comparisons (Appendix
\ref{appdx.prevcat}), we recover over 76\% of the
\citetalias{prochteretal06} absorbers, 80\% of
\citetalias{quideretal11}, and 72\% of
\citetalias{zhuandmenard13}.\footnote{\citetalias{nestoretal05} did
  not publish their line list so detailed comparison is not possible.}
We can also identify why we do not recover the remaining doublets,
with reasons ranging from the \citet{shenetal11} BAL QSO spectra were
not searched to the line was not automatically detected by our
algorithms in our normalized spectra. All catalogs are less complete
for weaker systems, and each catalog's unmatched sample largely has
$\EWlin{2796} \lesssim 1\Ang$.

Statistically and intuitively, disagreement at low equivalent
width---where all surveys become highly incomplete---is expected. For
example, we are 50\% complete at $\EWlin{2796} \approx 0.8\Ang$, while
\citetalias{zhuandmenard13} is roughly 60\% complete, so we should
only agree on $\approx\!30\%$ of doublets of this strength. There is a
large caveat because we all use SDSS spectra: our surveys are not
statistically independent. However, continuum placement factors
strongly into the automated detection of weak lines, which is why we
attempt to quantify this affect by refitting the continua in our
Monte-Carlo completeness tests.

In all cases, we detect \ion{Mg}{2} systems not in other
catalogs. Since \citetalias{prochteretal06} applied a hard $\EWlin{2796}
\ge 1\Ang$ cut, the vast majority of our unique systems are at lower
equivalent widths, but their redshift distribution typically follows
the full sample. The \citetalias{prochteretal06}-only sample 
favors higher redshifts where sky lines are abundant and make
verification difficult.

The redshifts of the \citetalias{quideretal11}- and
\citetalias{zhuandmenard13}-only samples are a fair sampling of the
full (parent) samples. However, our unique sample favors lower
redshift. The effect is more pronounced with respect to
\citetalias{zhuandmenard13}, which limited the search to
$\Delta z \ge 0.02$ red-ward of the quasar \ion{C}{4} emission line.

Another point of comparison is the $\EWlin{2796}/\EWlin{2803}$
ratio. For the matched samples, we tend to measure larger ratios than
\citetalias{quideretal11} and significantly larger than
\citetalias{zhuandmenard13}; \citetalias{prochteretal06} did not
publish \EWlin{2803}\ measurements. Our unique sample has a
significant number of systems with ratios less than unity, indicating we
identify more blended systems.

In Section \ref{sec.results}, we compare our science results with the
published catalogs, where such comparisons are suitable.

\begin{figure}[hbt]
  \begin{center}
    \includegraphics[width=0.49\textwidth]{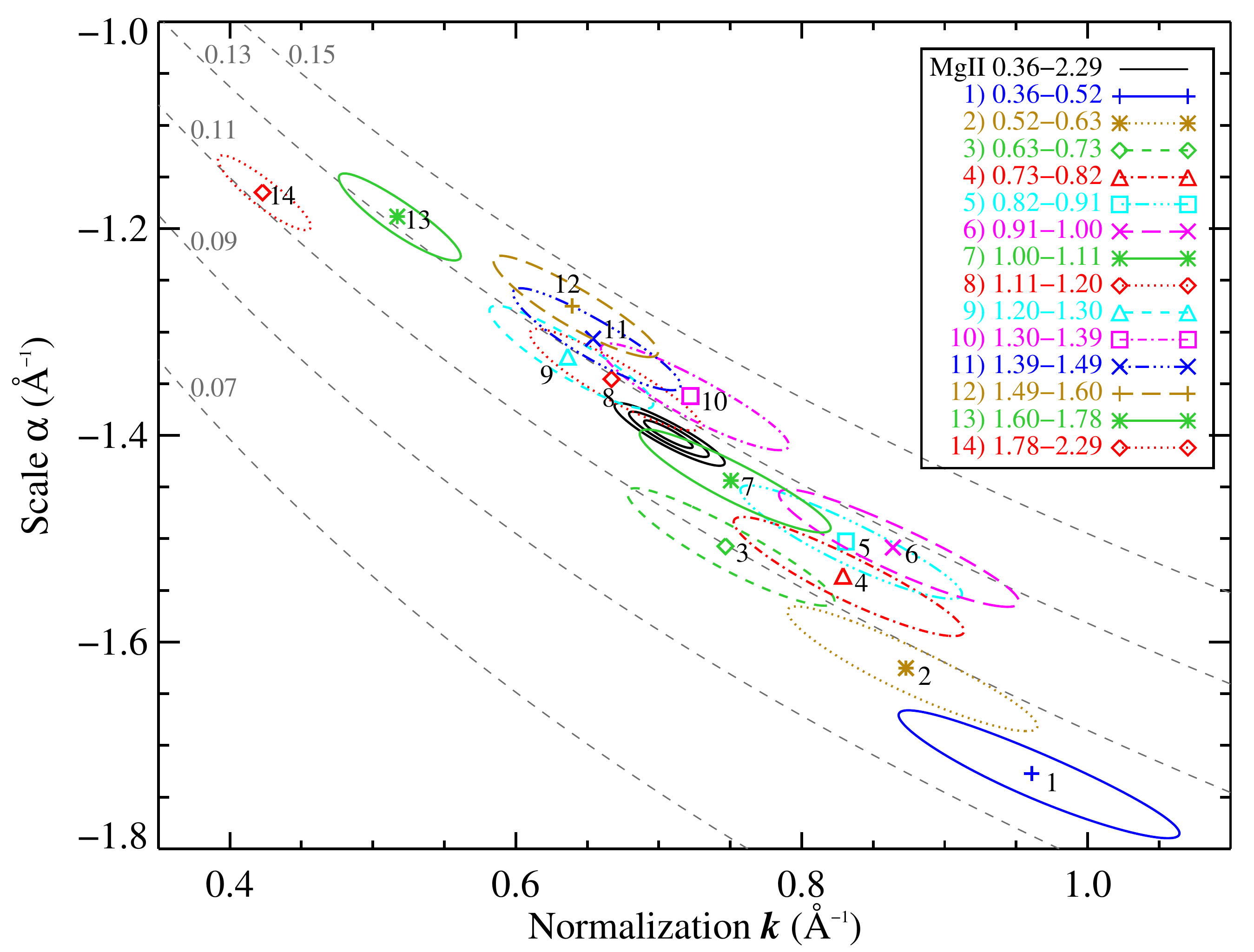}
  \end{center}
  \caption[Best-fit \ff{\EWr}\ parameters and errors.]
  {Best-fit \ff{\EWr}\ parameters and errors. We fitted the frequency
    distribution with an exponential function. The best-fit
    normalization $k$ and scale $\alpha$ and the 1-$\sigma$ error
    ellipses are plotted for the 14 small redshift bins (numbered
    points); the 1, 2, and 3-$\sigma$ contours are shown for the fit
    to the full sample (black ellipses). The best-fit parameters
    evolve fairly smoothly with redshift, as seen by comparing the
    ellipses with the constant \dNMgIIdX\ curves (gray, dashed lines;
    see Equation \ref{eqn.dndxfit}).
    \label{fig.fwellipse}
  }
\end{figure}

\section{Results}\label{sec.results}

Typically, we analyzed our \ion{Mg}{2} catalog as a whole
and in 14 small redshift bins with roughly 2500 doublets each. As
necessary, we modify the redshift binning to match other samples when
comparing to their results.

\subsection{Frequency Distribution}\label{subsec.freqdistr}

The equivalent-width frequency distribution \ff{\EWr}\ (sometimes
written, $d^{2}\Num_{\Mgp}/d \EWr/d X$) is the number of detections
$\Num_{\rm obs} (\EWr)$ per rest equivalent width bin $\Delta \EWr$,
per the total co-moving path length available, in the given
equivalent-width bin, $\Delta X(\EWr)$:
\begin{equation}
  \ff{\EWr} = \frac{\Num_{\rm obs}(\EWr)} {\Delta \EWr
    \,\DX{\EWr}}  \label{eqn.fewdef} {\rm ,}
\end{equation}
and it is $\Delta X(\EWr)$ that accounts for completeness. We modeled
\ff{\EWr}\ with an exponential, $\ff{\EWr} = k \exp(\alpha \EWr)$, and
fitted with the maximum likelihood method of \citet{cookseyetal10} for
$\EWlin{2796} \ge 0.8\Ang$, though the results are relatively
insensitive to a change of $\pm0.2\Ang$. The exponential model is a
very good description of the data (Figure \ref{fig.fw}), and the
best-fit parameters are given in Table \ref{tab.freqdistr}.

The frequency distribution flattens with increasing redshift, meaning
there are more strong absorbers relative to weak ones, from $\zmgii =
0.4 \rightarrow 2.3$. The relatively smooth and significant evolution
in the best-fit parameters from low-to-intermediate redshift can be
seen in Figure \ref{fig.fwellipse}.  We also show that the 
line density is lower at the extrema of the redshift range, where:
\begin{equation}
\frac{\displaystyle d \Num_{\Mgp}}{\displaystyle d X}\bigg |_{\rm fit} =
\frac{\displaystyle -k}{\displaystyle \alpha} e^{\alpha W_{\rm
    lim}} \label{eqn.dndxfit} {\rm ,}
\end{equation}
from the integral of the frequency distribution from a limiting
equivalent width, $W_{\lim}$, to infinity.

For \ion{Mg}{2}, the normalization ($k$) increases by a factor of
$\approx\!2.5$ from $z = 2.3 \rightarrow 0.4$, while the scale
($\alpha$) decreases by approximately 20\%. In comparison, for
\ion{C}{4}, $\alpha$ evolves little from $z = 4.5
\rightarrow 1.5$, while $k$ increases by three-fold, roughly
\citepalias{cookseyetal13}.

As in \citetalias{cookseyetal13}, we factor in the accepted
false-positive rate by scaling the original, measured frequency
distribution, $f_{0}(\EWr)$:
\begin{equation}
\ff{\EWr} = \bigg(1 - \frac{\displaystyle \dNFkIIdX}{\displaystyle
  \dNMgIIdX} \bigg) f_{0}(\EWr) \label{eqn.afpfw} {\rm ,}
\end{equation}
which results in an updated best-fit normalization of:
\begin{equation}
k = \bigg(1 - \frac{\displaystyle \dNFkIIdX}{\displaystyle
  (\dNMgIIdX)_{\rm fit}} \bigg) k_{0} {\rm .} \label{eqn.afpfwfit}
\end{equation}
In the latter equation, the denominator uses the integrated line
density from Equation \ref{eqn.dndxfit}. We report the propagated
errors in Table \ref{tab.freqdistr} and in the text.

From high-resolution, high-S/N spectroscopy of a smaller number of
quasars, \citet{churchilletal99a} and \citet{narayananetal07} both
measure a power-law \ff{\EWr}\ for $\EWlin{2796} \lesssim 0.3\Ang$
systems. SDSS---as an efficient, low-resolution, moderate-S/N
survey---naturally provides great statistics on the rare, strong
systems typically missing in smaller surveys. In
\citetalias{cookseyetal13}, the newly measured strong-end of the
\ion{C}{4} frequency distribution was also well-modeled by an
exponential, which provided the first detection of a break, since
previous (high-resolution, high-S/N, smaller sample) studies had
modeled the frequency distribution as a power-law.

\begin{deluxetable*}{ccccccccccc}
\tablewidth{0pc}
\tablecaption{\ion{Mg}{2} Results Summary \label{tab.freqdistr}}
\tabletypesize{\scriptsize}
\tablehead{ 
\colhead{(1)} & \colhead{(2)} & \colhead{(3)} & \colhead{(4)} & 
\colhead{(5)} & \colhead{(6)} & \colhead{(7)} & \colhead{(8)} & 
\colhead{(9)} & \colhead{(10)} & \colhead{(11)} 
\\ 
\colhead{$\langle z \rangle$} & \colhead{$z_{\rm lim}$} & 
\colhead{$\Num_{\rm obs}$} & \colhead{$\Delta X_{\rm max}$} & \colhead{$W_{50\%}$} & 
\colhead{\dNMgIIdz} & \colhead{\dNMgIIdX} & 
\colhead{$\Num_{\rm fit}$} & \colhead{$k$} & \colhead{$\alpha$} &  
\colhead{$\chi^{2}_{\rm red}$} \\ 
 & & 
 & & \colhead{(\AA)} 
 & & & 
 & \colhead{(\AA$^{-1}$)} & \colhead{(\AA$^{-1}$)} &  
}
\startdata
  1.10749 & $[  0.36539,   2.28259]$ &  34254 & 183071 &  0.81 & $0.293^{+0.002}_{-0.002}$ & $ 0.123^{+0.001}_{-0.001}$ &   22414 & $   0.71^{+   0.02}_{-   0.02}$ & $  -1.40^{+   0.01}_{-   0.01}$ &   0.535 \\ 
  0.45524 & $[  0.36539,   0.51988]$ &   2414 &  17037 &  1.04 & $0.162^{+0.004}_{-0.004}$ & $ 0.096^{+0.002}_{-0.002}$ &    1509 & $   0.96^{+   0.11}_{-   0.10}$ & $  -1.73^{+   0.06}_{-   0.06}$ &   1.519 \\ 
  0.57586 & $[  0.52003,   0.62992]$ &   2402 &  13739 &  0.82 & $0.191^{+0.005}_{-0.005}$ & $ 0.102^{+0.003}_{-0.003}$ &    1485 & $   0.87^{+   0.10}_{-   0.09}$ & $  -1.63^{+   0.06}_{-   0.06}$ &   2.422 \\ 
  0.68065 & $[  0.63003,   0.72993]$ &   2491 &  13048 &  0.74 & $0.220^{+0.006}_{-0.005}$ & $ 0.110^{+0.003}_{-0.003}$ &    1510 & $   0.75^{+   0.08}_{-   0.08}$ & $  -1.51^{+   0.06}_{-   0.06}$ &   1.327 \\ 
  0.77562 & $[  0.73005,   0.81998]$ &   2445 &  12057 &  0.72 & $0.251^{+0.006}_{-0.006}$ & $ 0.118^{+0.003}_{-0.003}$ &    1508 & $   0.83^{+   0.09}_{-   0.08}$ & $  -1.54^{+   0.06}_{-   0.06}$ &   1.617 \\ 
  0.86369 & $[  0.82002,   0.90992]$ &   2575 &  12320 &  0.71 & $0.274^{+0.007}_{-0.007}$ & $ 0.123^{+0.003}_{-0.003}$ &    1627 & $   0.83^{+   0.09}_{-   0.08}$ & $  -1.50^{+   0.05}_{-   0.05}$ &   1.240 \\ 
  0.94890 & $[  0.91002,   0.99998]$ &   2288 &  11944 &  0.80 & $0.295^{+0.007}_{-0.007}$ & $ 0.126^{+0.003}_{-0.003}$ &    1480 & $   0.86^{+   0.10}_{-   0.09}$ & $  -1.51^{+   0.06}_{-   0.06}$ &   1.658 \\ 
  1.04931 & $[  1.00003,   1.10999]$ &   2557 &  14924 &  0.93 & $0.292^{+0.006}_{-0.006}$ & $ 0.119^{+0.003}_{-0.003}$ &    1749 & $   0.75^{+   0.08}_{-   0.07}$ & $  -1.44^{+   0.05}_{-   0.05}$ &   1.535 \\ 
  1.15652 & $[  1.11001,   1.20000]$ &   2356 &  11850 &  0.76 & $0.332^{+0.008}_{-0.008}$ & $ 0.130^{+0.003}_{-0.003}$ &    1590 & $   0.67^{+   0.07}_{-   0.06}$ & $  -1.35^{+   0.05}_{-   0.05}$ &   1.836 \\ 
  1.25157 & $[  1.20005,   1.29999]$ &   2521 &  11904 &  0.70 & $0.330^{+0.008}_{-0.008}$ & $ 0.125^{+0.003}_{-0.003}$ &    1578 & $   0.64^{+   0.06}_{-   0.06}$ & $  -1.32^{+   0.05}_{-   0.05}$ &   1.060 \\ 
  1.34541 & $[  1.30009,   1.38999]$ &   2515 &  10176 &  0.53 & $0.369^{+0.010}_{-0.009}$ & $ 0.135^{+0.004}_{-0.003}$ &    1550 & $   0.72^{+   0.07}_{-   0.07}$ & $  -1.36^{+   0.05}_{-   0.05}$ &   1.237 \\ 
  1.43499 & $[  1.39003,   1.48998]$ &   2431 &  10394 &  0.58 & $0.374^{+0.010}_{-0.010}$ & $ 0.133^{+0.003}_{-0.003}$ &    1543 & $   0.65^{+   0.06}_{-   0.06}$ & $  -1.31^{+   0.05}_{-   0.05}$ &   2.267 \\ 
  1.54058 & $[  1.49024,   1.59994]$ &   2321 &  10060 &  0.63 & $0.400^{+0.010}_{-0.010}$ & $ 0.138^{+0.004}_{-0.003}$ &    1527 & $   0.64^{+   0.06}_{-   0.06}$ & $  -1.27^{+   0.05}_{-   0.05}$ &   1.652 \\ 
  1.67796 & $[  1.60001,   1.77974]$ &   2467 &  13671 &  0.82 & $0.398^{+0.009}_{-0.009}$ & $ 0.132^{+0.003}_{-0.003}$ &    1771 & $   0.52^{+   0.05}_{-   0.04}$ & $  -1.19^{+   0.04}_{-   0.04}$ &   1.064 \\ 
  1.91822 & $[  1.78010,   2.28259]$ &   2471 &  23091 &  1.22 & $0.360^{+0.007}_{-0.007}$ & $ 0.111^{+0.002}_{-0.002}$ &    1987 & $   0.42^{+   0.04}_{-   0.04}$ & $  -1.16^{+   0.04}_{-   0.04}$ &   2.489
\enddata
\tablecomments{
Summary of the most common redshift bins and data used for the various analyses.
Columns 1--2 give the median, minimum, and maximum redshifts for the observed number of doublets (Column 3), and the maximum co-moving pathlength in the redshift bin is given in Column 4.
The 50\% completeness limit from the Monte Carlo tests is in Columns 5.
The redshift and co-moving absorber line densities for $\EWr \ge 1.0\Ang$ are in Columns 6--7.
The frequency distribution was fit with an exponential $\ff{\EWr} = k \exp(\alpha\EWr)$ for $\Num_{\rm fit}$ absorbers with $\EWr \ge 0.8\Ang$ (Column 8), and the best-fit parameters are given in Columns 9--10.
The reduced $\chi^{2}$ from the best fit and \ff{\EWr}\ (in bins with $\approx100$ doublets each) is given in Column 11.
}
\end{deluxetable*}

\subsection{\ion{Mg}{2} Absorber Line Density}\label{subsec.dndx}

The absorber line density is the completeness-corrected number of
\ion{Mg}{2} doublets within the given \EWlin{2796} limits, normalized
by the total redshift or co-moving path length available, i.e.,
\dNMgIIdz\ or \dNMgIIdX, respectively.  We subtract the accepted
false-positive line density, \dNFkIIdz\ or \dNFkIIdX, from our quoted
\ion{Mg}{2} line densities, for the appropriate redshift bin (Figure
\ref{fig.userbias}). For $\EWr \ge 1\Ang$, $\dNFkIIdX =
0.0021^{+0.0001}_{-0.0001}$ ($0.3 < z < 2.3$);
$0.0015^{+0.0003}_{-0.0003}$ ($z < 0.63$);
$0.0020^{+0.0001}_{-0.0001}$ ($0.63 \ge z < 1.78$); and
$0.0045^{+0.0011}_{-0.0010}$ ($z \ge 1.78$).

\begin{figure}[hbt]
  \begin{center}
    \includegraphics[width=0.49\textwidth]{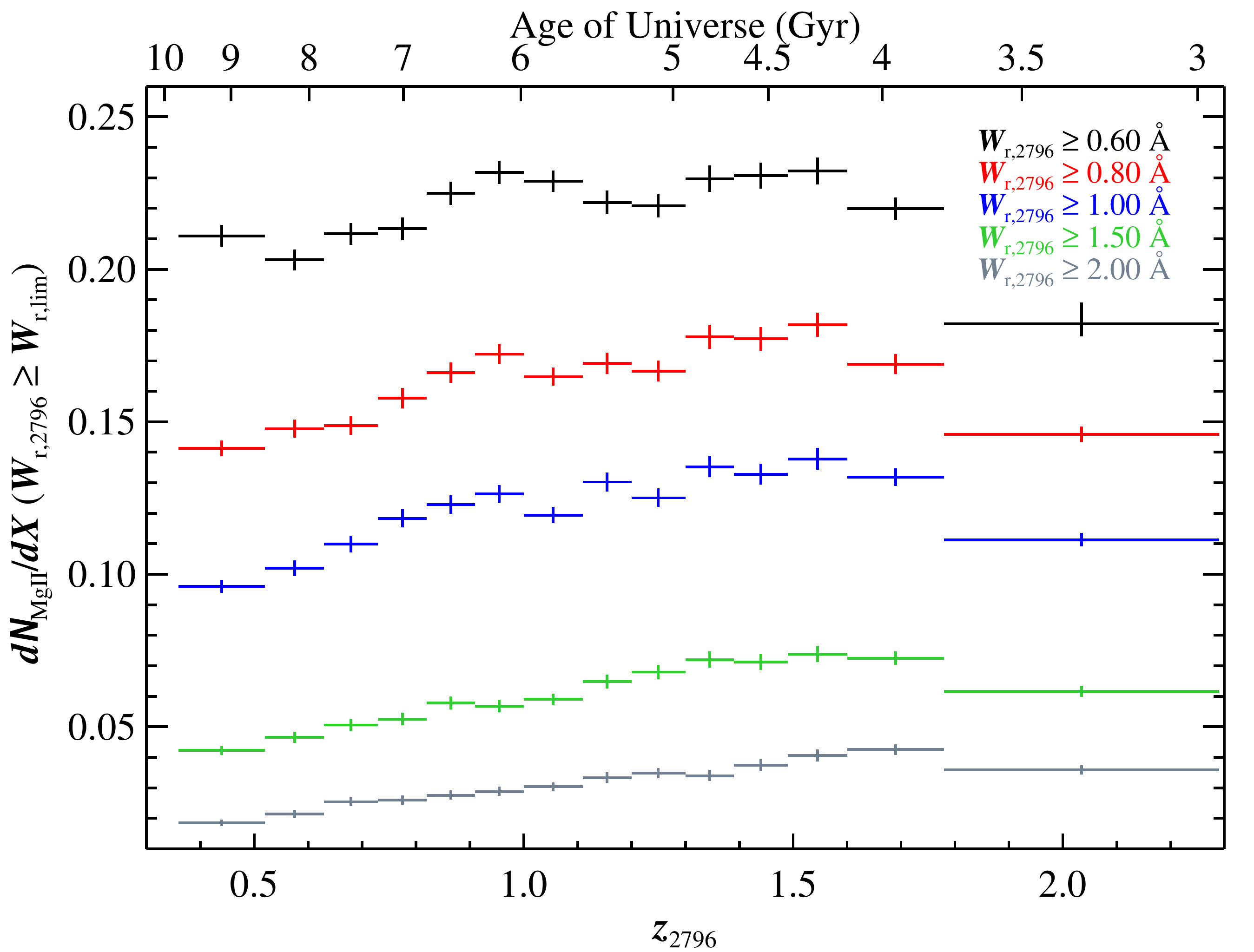}
  \end{center}
  \caption[\ion{Mg}{2} absorber line density evolution.]
  {\ion{Mg}{2} absorber line density evolution. The number of
    absorbers per co-moving path length increases from $\zmgii = 0.4
    \rightarrow\ \approx\!1.6$.  As expected from the flattening of
    \ff{\EWr}\ towards higher redshift (Figure \ref{fig.fw}), the
    stronger systems show more evolution, increasing in number towards
    higher redshift. We are typically 50\% complete at $\EWr \approx
    0.8\Ang$.
    \label{fig.dndx}
  }
\end{figure}

In Figure \ref{fig.dndx}, we compare \dNMgIIdX\ for a range of
limiting equivalent widths, \EWlin{lim}. There is significant
differential evolution based on \EWlin{lim}, as predicted by the
changing shape of \ff{\EWr} over time (Figure \ref{fig.fw}). The $\EWr
\ge 1\Ang$ absorbers increase by approximately 45\% from $z = 0.4
\rightarrow 1.5$, while the $\EWr \ge 2\Ang$ systems increase by a a
factor of 2.3 in over the same interval.

The peak in \dNMgIIdX\ appears to shift to higher redshift for larger
\EWlin{lim}, from $z \approx 1.5$ ($\EWr \ge 0.8\Ang$) to $z \approx
1.7$ ($\EWr \ge 2\Ang$). The cosmic star-formation rate density
($\rho_{\ast}$) peaks at $2 \lesssim z \lesssim 3$
\citep{bouwensetal10}. Observations have associated strong \ion{Mg}{2}
systems with outflows from star-forming galaxies. It is reasonable
that there would be more strong systems closer to the peak in
$\rho_{\ast}$. The changing shape of \ff{\EWr}\ supports this result.

\citet{matejekandsimcoe12} showed that weaker, $0.3\Ang \le
\EWlin{2796} < 1\Ang$ systems evolve surprisingly little from $z
\approx 5.5 \rightarrow 0.4$. This persistent, weaker \ion{Mg}{2}
population arises either from pre-enrichment at early times or
constant replenishment (at a density-preserving rate), for most of the
lifetime of the universe.


We note that the 50\% completeness limit in our highest redshift bin
is $\EWlin{2796} \approx 1.2\Ang$, significantly higher than our other
bins (due to poor skyline subtraction). We extend our analysis in this
bin to below this value in order to compare to results using the
canonical $1\Ang$ limit. As seen in Figure \ref{fig.dndx}, \dNMgIIdX\
does consistently decrease for equivalent-width cuts greater than
the 50\% completeness limit, so the turnover is likely real. We
discuss the possible physical explanations for the observed evolution
in \dNMgIIdX\ in Section \ref{sec.discuss}.

\begin{figure*}[hbt]
  \begin{center}$
    \begin{array}{cc}
      \includegraphics[width=0.49\textwidth]{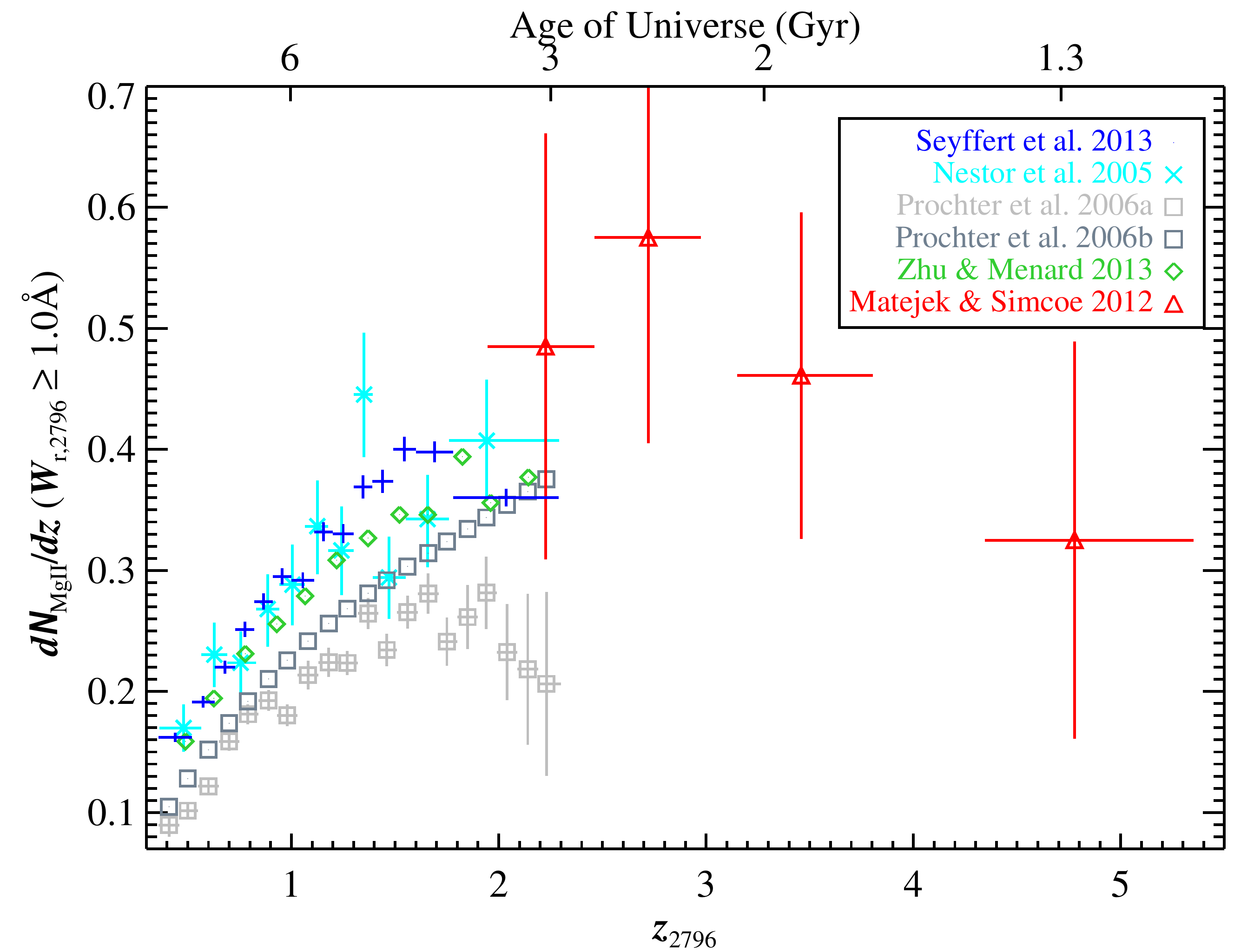}  & 
      \includegraphics[width=0.49\textwidth]{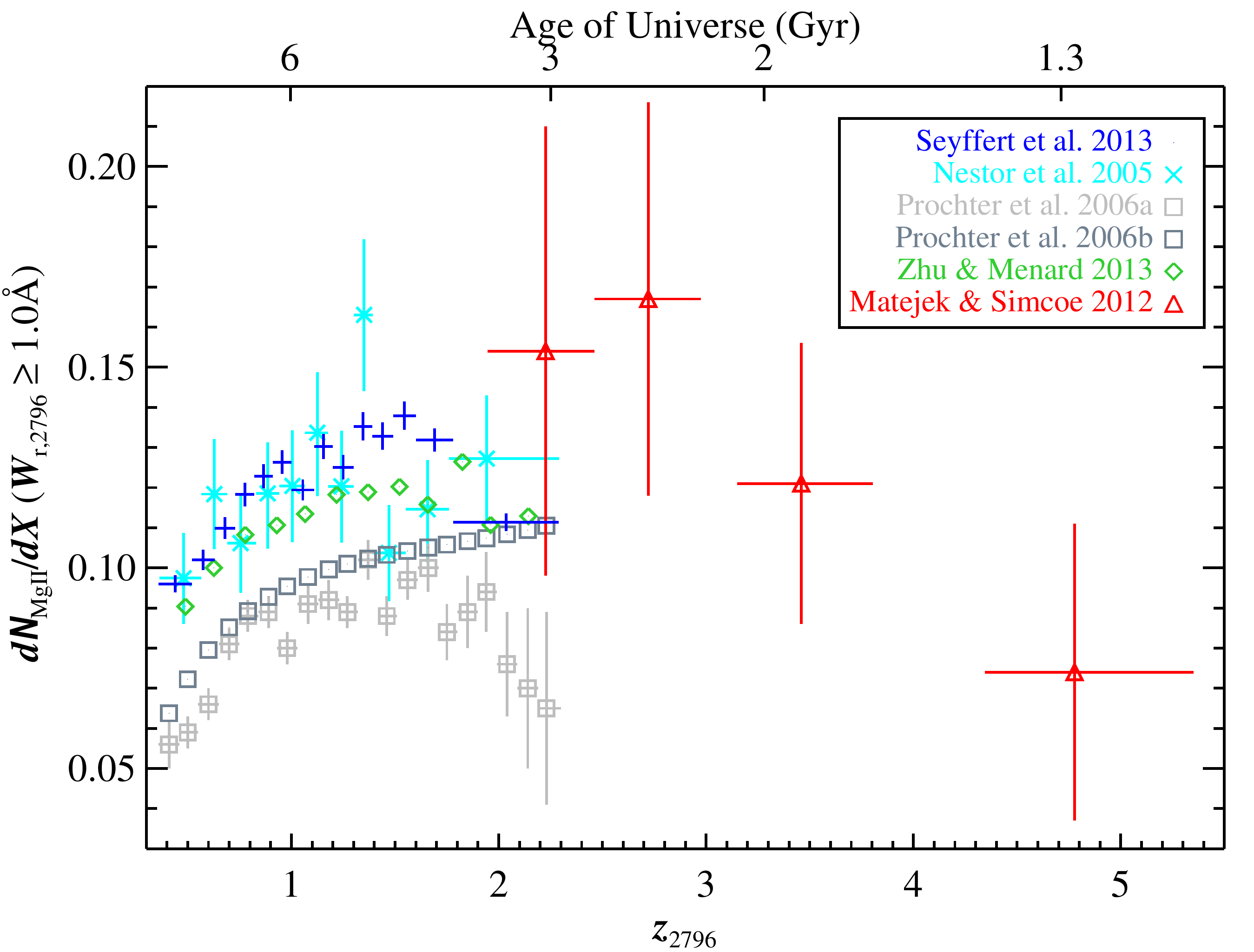} 
    \end{array}
    $\end{center}
  \caption[Redshift and co-moving path length \ion{Mg}{2} absorber line
  densities.]
  {Redshift (left) and co-moving path length (right) \ion{Mg}{2} line
    densities. We compare apples-to-apples \dNMgIIdz\ with $\EWr \ge
    1\Ang$ from six surveys: this work (blue plusses);
    \citetalias{nestoretal05} (cyan crosses);
    \citetalias{prochteretal06} (light gray squares); \citet[][dark
    gray squares]{prochteretal06grb}; \citetalias{zhuandmenard13}
    (green diamonds); and \citet[red
    triangles]{matejekandsimcoe12}. The redshift density steadily
    increases from $z \approx 0.4$ to $\approx1.75$.  The relatively
    flat \dNMgIIdX\ indicates modest evolution in the product of the
    co-moving number density and the physical cross-section of the
    absorbing clouds.
    \label{fig.dndzlit}
  }
\end{figure*}

Quasars are not the only background source suitable for
absorption-line studies; gamma-ray bursts (GRBs) have also been used
to study intergalactic \ion{Mg}{2} absorption systems. In the seminal
study on \ion{Mg}{2} doublets in GRB sightlines,
\citet{prochteretal06grb} identified roughly four-times as many
systems as would be expected based on the \citetalias{prochteretal06}
quasar \dNMgIIdX\ measurement. The authors discussed possible
explanations: high-velocity, intrinsic \ion{Mg}{2} population in GRB
hosts; bias due to dust obscuration in quasar sightlines; GRBs having
larger cross-sections; and effects of gravitational lensing. However,
subsequent studies never fully resolved the difference between GRB and
QSO \dNMgIIdX. Recently, \citet{cucchiaraetal13} tackled the issue
by increasing the GRB statistics---including a completely independent
sample from \citet{prochteretal06grb}. Using the
\citetalias{zhuandmenard13} \dNMgIIdX\ values, which, as seen in
Figure \ref{fig.dndzlit}, are larger than \citetalias{prochteretal06},
\citet{cucchiaraetal13} concluded that there is no statistical
difference between GRB and QSO sightlines. The latest \dNMgIIdz\
results from this work are $\approx\!10\%$--20\% larger than those
from \citetalias{zhuandmenard13}, which brings the GRB and QSO results
further in to agreement.

\subsection{Comparing with Previous SDSS \ion{Mg}{2}
  Results}\label{subsec.prevrslt}

Previous SDSS \ion{Mg}{2} surveys typically fitted \ff{\EWlin{2796}}\
with an exponential model and measure \dNMgIIdz\ as a function of
redshift, and here we compare their results with ours.  The biggest
caveat when comparing results is: all SDSS \ion{Mg}{2} studies are
{\it statistically} correlated, since they are based on subsets of the
same observations. In addition, there are dependencies between
methodologies. Since \citetalias{zhuandmenard13} used
\citetalias{quideretal11} as a training set, and the latter, in turn,
was based heavily on the methodologies of \citetalias{nestoretal05},
the results of these studies are highly correlated.

All studies that fitted \ff{\EWr}\ \citepalias{nestoretal05,
  zhuandmenard13} find it well-modeled by an exponential, with no
apparent break over $0.3\Ang \lesssim \EWlin{2796} \lesssim 10\Ang$.
These studies fitted the frequency distribution as: $\ff{\EWr} =
d\Num_{\Mgp}/d\EWr/dz = \Num^{\ast}/W^{\ast} \exp(-\EWr/W^{\ast})$,
which relates better to a Schechter function formalism. Relating our
fit parameters to this formalism yields: $k = -\Num^{\ast}\alpha$ and
$\alpha = -(W^{\ast})^{-1}$. \citetalias{nestoretal05} found that the
characteristic equivalent width, $W^{\ast}$, increased steadily in
their three redshift bins covering $z = 0.4 \rightarrow 2.3$. Over
their 12 redshift bins, \citetalias{zhuandmenard13} measured a
turnover in $W^{\ast}$ at $z \approx 1.75$. \citet{matejekandsimcoe12}
measured a monotonically decreasing $W^{\ast}$ from $z = 2 \rightarrow
5.5$, which mapped well on to the \citetalias{nestoretal05} values.

For comparison, we fit the frequency distribution in redshift space
and convert the best-fit parameters, $k$ and $\alpha$, to the fairly
comparable quantities, $\Num^{\ast}$ and $W^{\ast}$. We measure
steadily increasing $W^{\ast}$ with redshift, in agreement with
\citetalias{nestoretal05}.\footnote{Since our highest redshift bin
  begins where \citetalias{zhuandmenard13} detect a turnover, we fitted
  \ff{\EWr}\ in three bins matching theirs and still detect a
  monotonically increasing $W^{\ast}$.}  \citetalias{nestoretal05}
detect no evolution in $\Num^{\ast}$ over $z = 0.4$ to 2.3 (three
redshift bins). \citet{matejekandsimcoe12} also detect no evolution
from $z = 2$ to 5.5 (also three bins). However, the
\citeauthor{matejekandsimcoe12} values are about 50\% larger than the
\citetalias{nestoretal05} measurements, with no obvious evolution in
the total of six redshift bins to bring this about. On the other hand,
$\Num^{\ast}$ as estimated by \citetalias{zhuandmenard13} steadily
increases from $z = 0.4$ to 2.3, mapping well on to the high-redshift
values.

However, as shown in Figure \ref{fig.fwellipse}, the fit parameters
are correlated: small increases in $W^{\ast}$ (or decreases in $\alpha$)
decreases $\Num^{\ast}$ (increases $k$). The measured \dNMgIIdz,
which is a well-measured quantity, basically defines the error
ellipse. Therefore, though \citetalias{nestoretal05} and our
$W^{\ast}$ values evolve consistently on to the high-redshift
measurements, both our $\Num^{\ast}$ values disagree with the
high-redshift estimates. For us, this manifests as a turnover in
$\Num^{\ast}$ at $z \approx 1.5$, while the
\citetalias{zhuandmenard13} values steadily increase.

We compare our $\EWlin{2796} \ge 1\Ang$ \dNMgIIdz\ and \dNMgIIdX\
values to the other studies in Figure \ref{fig.dndzlit}.  We compiled
\dNMgIIdz\ from the literature as follows:
\citetalias{nestoretal05}---Figure 9;
\citet{prochteretal06grb}---best-fit polynomial, updating
\citetalias{prochteretal06};\footnote{\citet{prochteretal06grb}
  updated the \citetalias{prochteretal06} analysis, increasing their
  \ion{Mg}{2} line density by $\approx\!20\%$.}
\citetalias{zhuandmenard13}---Figure 13; and
\citet{matejekandsimcoe12}---Table 5. As for \dNMgIIdX, we use:
\citetalias{prochteretal06}---Table 3 and
\citet{matejekandsimcoe12}---Table 5. As needed, we convert one line
density to the other using $dz/dX$ or its inverse,\footnote{See
  footnote \footnotemark[\ref{fn.dxdz}].} computed at the appropriate
redshift.

For all studies, the redshift density steadily increases from $z
\approx 0.4$ to at least $z \approx 1.8$, and comparing to the
high-redshift values \citep{matejekandsimcoe12}, there appears to be a
peak between $z \approx 1.5$ and 3. 

However, expansion of the universe contributes to the evolution of
\dNMgIIdz. Thus, we turn to \dNMgIIdX, where the normalization by
co-moving path length removes the effect of passive evolution. In
Figure \ref{fig.dndzlit}, \dNMgIIdX\ evolves less strongly. Examining
just the SDSS results, the peak appears to be between $z \approx 1.5$
and 2. The $z = 2$ to 3 values from \citet{matejekandsimcoe12} are
larger than the highest-redshift SDSS values but consistent within the
large uncertainties.

Now we discuss why the SDSS \dNMgIIdX\ measurements, which are drawn
from the same actual spectra, differ by amounts ranging from
$\approx\!15\%$ (i.e., ours relative to \citetalias{zhuandmenard13})
to $\approx\!50\%$ (relative to \citetalias{prochteretal06}). Since
the formal errors quoted by surveys ($\approx\!1\%$) are much smaller
than the differences, we conclude that the uncertainties are
dominated by systematic effects explored below, in addition to
completeness corrections, which are not discussed. Two systematic effects
contribute to \dNMgIIdX\ differences: Malmquist bias and variations
in \EWlin{2796}\ measurements.  The Malmquist bias refers to how more
absorbers scatter to above $\EWlin{lim}$ than scatter to below, due to
the \EWr\ uncertainties, for distributions steeply rising
toward weak absorbers.

Due to the exponential nature of \ff{\EWr}, a small change in
\EWlin{lim}\ can have significant impact on the quoted \dNMgIIdX\ (see
Equation \ref{eqn.dndxfit}); also, systematic differences in
equivalent-width measurements affect completeness corrections,
basically shifting a completeness curve to higher or lower \EWr. Since
there is an offset in the relative equivalent-width ``zero point''
between studies (see Appendix \ref{appdx.prevcat}), our
equivalent-width {\it limits} are effectively different and cause a
shift in \dNMgIIdX. For example, the median deviation of
$\EWlin{ZM13}-\EWlin{S13}$ for matched absorbers within
$\pm1\sigma_{W}$ of $\EWlin{2796} = 1\Ang$ (where it matters most) is
$\approx\!0.01\Ang$ to $0.02\Ang$ but with large scatter (median
absolute deviation of
$\approx\!0.15\Ang$). 
If we compare our \dNMgIIdX\ values for $\EWlin{2796} \ge 1.1\Ang$ to
the \citetalias{zhuandmenard13} values for $\EWlin{2796} \ge 1\Ang$ at
$z < 1.75$, they agree exceedingly well.

\citetalias{prochteretal06} had \EWlin{2796} measurements that
differed substantially from \citetalias{quideretal11},
\citetalias{zhuandmenard13}, and our values. Hence, the
\citetalias{prochteretal06} suffer significantly from the relative
\EWr\ ``zero point'' issue. This effect, some unknown Malmquist bias,
and, most importantly, different completeness corrections lead to
\citetalias{prochteretal06} differing the most from our and other
surveys.

Ultimately, the points of consensus for SDSS \ion{Mg}{2} systems are:
\ff{\EWlin{2796}}\ is well-modeled by an exponential; \dNMgIIdX\ peaks
at $z \approx 1.5$; and the magnitude of \dNMgIIdX\ is largely
consistent with the EDR results \citepalias{nestoretal05}, with which
\citet[][partial DR5]{lundgrenetal09} also agree.

\begin{figure}[hbt]
  \begin{center}
    \includegraphics[width=0.49\textwidth]{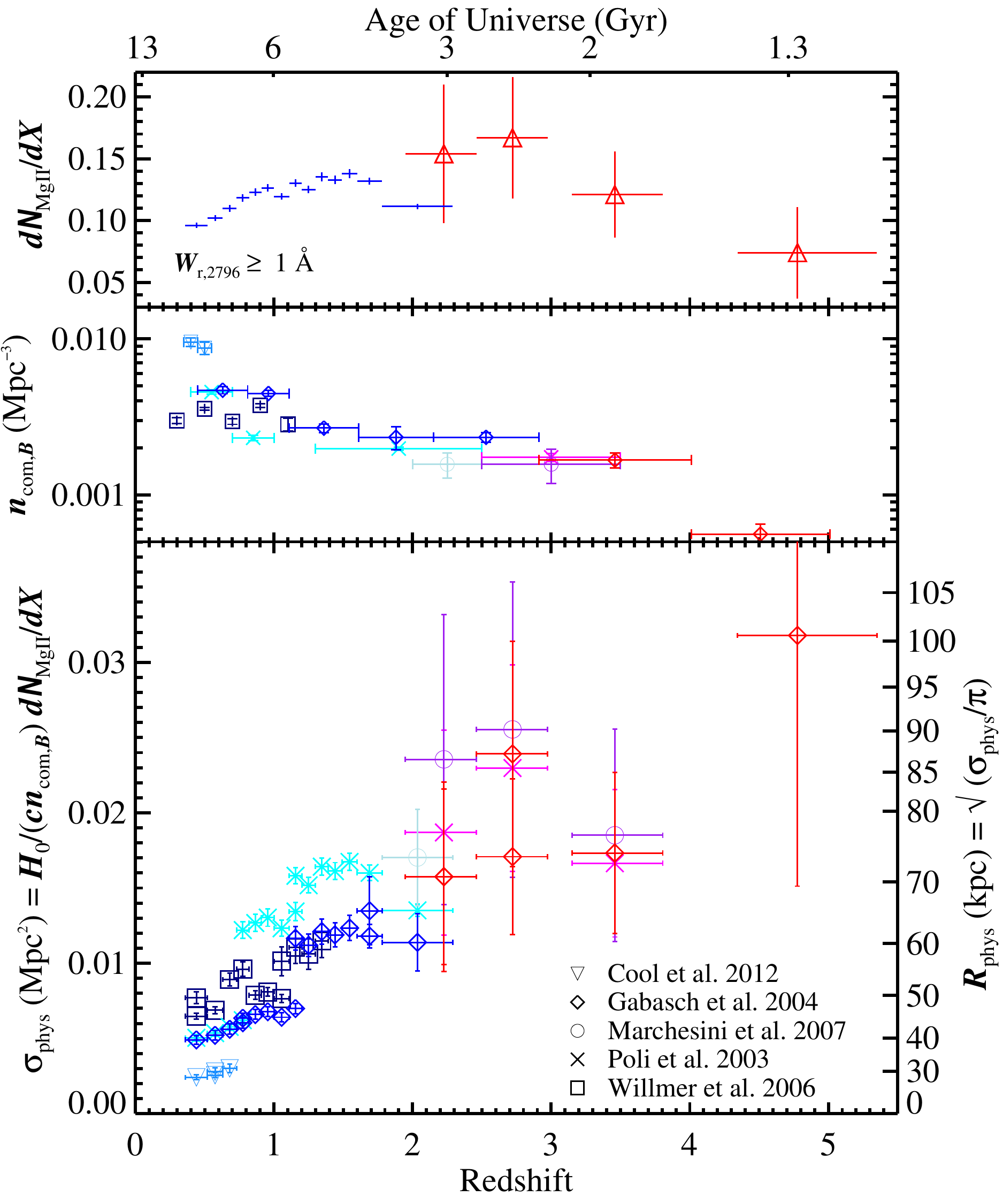}
  \end{center}
  \caption[\ion{Mg}{2}-absorbing halo cross-section estimate.]
  {\ion{Mg}{2}-absorbing halo cross-section estimate. Assuming that
    all $\EWlin{2796} \ge 1\Ang$ doublets are in {\it B}-band-selected
    galaxy halos (with $L \ge 0.5\,L^{\ast}$), we estimated the
    galaxy-\ion{Mg}{2} cross-section as a function of \zmgii\
    (Equation \ref{eqn.dndxphys}). {\it Top}: We used \dNMgIIdX\ from
    the current study (blue plusses) and \citet[][red
    triangles]{matejekandsimcoe12}. {\it Middle}: We calculated the
    co-moving number density of galaxies by integrating the $B$-band
    luminosity functions from the listed studies, down to
    $0.5\,L^{\ast}$. The galaxies are bright, typically star-forming
    galaxies, known to be associated with $z \lesssim 2$ \ion{Mg}{2}
    absorbers \citep[e.g.,][]{martinandbouche09, rubinetal13}. {\it
      Bottom}: The inferred cross-sections from the SDSS sample are
    comparable to observations \citep[e.g.,][]{chenetal10a,
      bordoloietal11, nielsenetal13}. The various studies providing
    the luminosity function parameters are linked by symbol across
    redshift and by color between the middle and bottom panels. If
    possible, we show multiple \ncomb\ and \sigphys\ per redshift,
    because the scatter between different studies characterizes
    the uncertainty in \sigphys\ better than the estimated errors of
    any one survey.
    \label{fig.crosssec}
  }
\end{figure}

\section{Discussion}\label{sec.discuss}

Our \ion{Mg}{2} catalog, in conjunction with the high-redshift survey
of \citet{matejekandsimcoe12}, traces the cosmic chemical enrichment
cycle from $\zmgii = 0.4 \rightarrow 5.5$, or from 10\,Gyr to 1\,Gyr
after the Big Bang. Here we examine how the evolution in \dNMgIIdX\
relates to galaxy evolution and how various absorber populations evolve.

\subsection{\ion{Mg}{2} Evolution}\label{subsec.mgiievo}

Considering the direct evidence for strong \ion{Mg}{2} systems arising
in the gaseous halos of galaxies \citep[e.g.,][]{chenetal10a,
  churchilletal13}, especially star-forming ones
\citep[e.g.,][]{martinandbouche09, rubinetal13}, we estimate the
physical cross-section of \ion{Mg}{2} absorbers, \sigphys, by assuming
the co-moving number density of clouds equals \ncomb, the co-moving
number density of $B$-band-selected galaxies, i.e.,:
\begin{equation}
\frac{\displaystyle d \Num_{\Mgp}}{\displaystyle d X} =
\frac{\displaystyle c}{\displaystyle H_{0}} \ncomb \sigphys {\rm
  .} \label{eqn.dndxphys}  
\end{equation}
These galaxies are bright, typically star-forming
galaxies, known to be associated with $z \lesssim 2$ \ion{Mg}{2}
absorbers \citep[e.g.,][]{martinandbouche09, rubinetal13}.

In Figure \ref{fig.crosssec}, we calculate \ncomb\ by integrating the
$B$-band luminosity functions from \citet[][crosses]{polietal03},
\citet[][diamonds]{gabaschetal04}, \citet[][squares]{willmeretal06},
\citet[][circles]{marchesinietal07}, and \citet[][inverted
triangles]{cooletal12}, down to $0.5\,L^{\ast}$. At $\zmgii < 2.3$,
where the cross-section uncertainties are dominated by the \ncomb\
errors, $\sigphys\ \approx 0.005\Mpc^{2}$ to $0.015\Mpc^{2}$. Assuming
the \ion{Mg}{2}-absorbing gas is distributed uniformly in a projected
disk on the sky, the radius would be 40\kpc\ to 70\kpc\ (right-hand
axis) for the SDSS redshift range, which agrees well with the observed
radial profile of \ion{Mg}{2}-absorbing galaxy halos \citep[][though
see \citealt{werketal13}]{chenetal10a, bordoloietal11,
  nielsenetal13} and permits variation in the luminosity limit
and\slash or covering fraction. The cross-section is larger at higher
redshifts, suggesting that $L < 0.5\,L^{\ast}$ galaxies may
contribute.

However, ultra-strong ($\EWr \ge 3\Ang$) \ion{Mg}{2} systems might be
associated with galaxy group gas \citep{gauthier13}. If
we limit \dNMgIIdX\ to $1\Ang \le \EWr < 3\Ang$ systems, the
cross-sections is reduced by $5\%$ to $10\%$ from $z = 0.4$ to
2.3. Recent work by \citet{werketal13} on the CGM of $z \approx 0.2$,
$L^{\ast}$ galaxies showed that the covering fraction for $\EWlin{2796}
\ge 1\Ang$ is small ($\lesssim 30\%$ within 50\kpc). It is difficult
to compare the SDSS systems with those of \citet{werketal13} because
they have small statistics on the $\EWr \ge 1\Ang$ absorbers.

\citet{chenetal10b} find a tighter correlation between \EWlin{2796}\
and the projected distance to the host galaxy if the latter is scaled
by the $B$-band luminosity: $R_{\rm gas} = 75 \times
(L_{B}/L^{\ast})^{0.35}\,h^{-1}\kpc$. If we adopt this model, we can
estimate the limiting luminosity, $L_{\rm lim}$, across redshift that
best reproduces the observed \dNMgIIdX. Roughly, $L_{\rm lim} \approx
0.015\,L^{\ast}$ at $\zmgii = 0.4$ and increases to
$\approx0.1\,L^{\ast}$ at $\zmgii = 1.8$ before dropping slightly in
the last SDSS redshift bin. This would increase \ncomb\ by up to a
factor of two, but both \ncomb\ and an estimated \sigphys\ would
evolve roughly as shown. The big caveat to applying the
\citet{chenetal10b} relation is that it was calibrated at $\zmgii <
0.5$, for $\EWlin{2796} \lesssim 3\Ang$, and with $L \gtrsim
0.1\,L^{\ast}$ galaxies.

In \citetalias{cookseyetal13}, the \ion{C}{4}-absorbing cross-section
was estimated to be roughly constant from $z_{1548} \approx 1.5
\rightarrow 4.5$, assuming the co-moving density of clouds equals that
of UV-selected galaxies, and the evolution of $n_{\rm com,UV}$ drives
the approximately two-fold decrease of $d\Num_{\rm C\,IV}/dX$ in the
redshift interval. In comparison, evolution in \dNMgIIdX\ might be
driven by the (large) increase in the \ion{Mg}{2}-absorbing
cross-section of galaxies from low-to-high redshift (Figure
\ref{fig.crosssec}). Though, the evolution might be a result of a
changing population of galaxies (e.g., $L < 0.5\,L^{\ast}$) that host
\ion{Mg}{2} absorbers. We will explore these scenarios in a future
paper.


\begin{figure}[hbt]
  \begin{center}
    \includegraphics[width=0.49\textwidth]{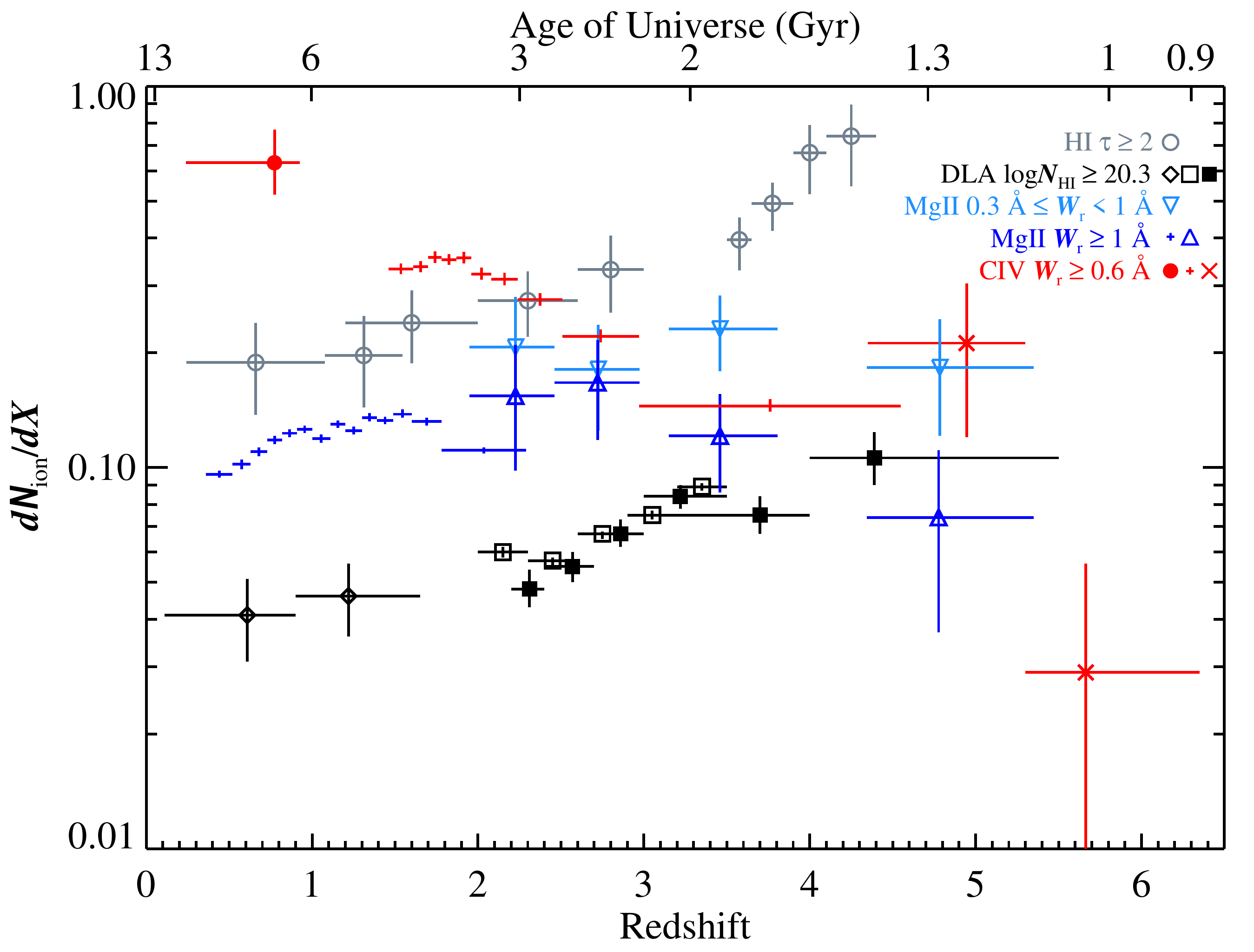}
  \end{center}
  \caption[Line density evolution of various absorber populations.]
  {Line density evolution of various absorber populations: $\tau > 2$
    \ion{H}{1} systems (i.e., LLS and DLA; gray); DLAs only (black);
    strong \ion{Mg}{2} and \ion{C}{4} systems (blue and red,
    respectively); and weaker \ion{Mg}{2} systems at high redshift
    (cyan) from the following studies:
    \ion{H}{1}---\citet[][circles]{fumagallietal13};
    DLA---\citet[][diamonds]{raoetal06},
    \citet[][filled squares]{prochaskaandwolfe09},
    \citet[][squares]{noterdaemeetal12}; \ion{Mg}{2}---current work
    (plusses), \citet[][triangles]{matejekandsimcoe12}; and
    \ion{C}{4}---\citet[][filled circle]{cookseyetal10},
    \citetalias{cookseyetal13} (plusses),
    \citet[][crosses]{simcoeetal11}.
    \label{fig.dndxions}
  }
\end{figure}

\subsection{Evolution of Various Absorber
  Populations}\label{subsec.multiion}

Most, if not all, SDSS \ion{Mg}{2} systems have other metal lines and,
where coverage exists, \ion{H}{1} absorption.  These associated
transitions assisted our visual verification, and in their automated
procedure, \citetalias{zhuandmenard13} used detection of (probable)
\ion{Fe}{2} lines to disentangle \ion{Mg}{2} doublets from other
possible identifications. In addition to transitions of \ion{H}{1}
Lyman series and iron, \ion{Mg}{2} systems can have aluminium,
silicon, and\slash or carbon absorption. Here we examine \dNdX{ion}\
of various samples, selected on the basis of different
transitions---DLAs, Lyman-limit systems (LLSs), strong \ion{C}{4}, and
strong and weaker \ion{Mg}{2}---and discuss the evolution of these
common tracers of the CGM and IGM across cosmic time (Figure
\ref{fig.dndxions}).

The differently evolving absorbers in Figure \ref{fig.dndxions} could
be physically explained, independently, as follows: (i) optical depth
$\tau \ge 2$ \ion{H}{1} systems decrease with time as the universe
becomes more ionized; (ii) strong \ion{C}{4} systems increase with the
increasing metallicity of the universe; (iii) strong \ion{Mg}{2}
systems evolve with time in lock-step with $\rho_{\ast}$; and (iv)
weaker \ion{Mg}{2} systems are established early or constantly
replenished so as to evolve little. Though these explanations may be
partially or completely true, they neither consider nor shed light on
the evolution of the other ions. We attempt to describe the evolution
of common QAL systems holistically.

We postulate that each ion has multiple sub-populations, which
together, yield the observed \dNdX{ion}. A difficulty arises in how to
fairly compare the data, considering the varying equivalent-width,
column-density, or optical-depth limits. For this discussion, we
simply compile the best published values for a given cutoff. For $\tau
\ge 2$ \ion{H}{1} systems (i.e., LLSs and DLAs), we use the
compilation by \citet{fumagallietal13}, which is based on
their measurement at $z = 2.8$ and \citet[][values centered at $z <
1.5$]{ribaudoetal11a}, \citet[][$1.5 < z < 3$]{omearaetal13}, and
\citet[][$z > 3.5$]{prochaskaetal10}. The DLA-only $d\Num/dX$ values
are from \citet[][$z < 2$]{raoetal06}, \citet[][$2 < z <
3.5$]{noterdaemeetal12}, and \citet[][$2 < z <
5.5$]{prochaskaandwolfe09}. The current work and \citet[][$z >
2$]{matejekandsimcoe12} provide \dNMgIIdX\ for $\EWlin{2796} \ge
1\Ang$, and the line densities for $0.3\Ang \le \EWlin{2796} < 1\Ang$
absorbers are from the latter survey. For $\EWlin{1548} \ge 0.6\Ang$
\ion{C}{4} systems, we use \citet[][$z < 1$]{cookseyetal10},
\citetalias{cookseyetal13} ($1.5 < z < 4.5$), and \citet[][$z >
4.5$]{simcoeetal11}.

In the original SDSS \ion{Mg}{2} paper, \citetalias{nestoretal05}
discussed ``multiple \ion{Mg}{2}'' populations. They noted that DLAs,
possible disks of star-forming galaxies, are a fraction of strong
\ion{Mg}{2} systems and that bright, spiral galaxies are not
guaranteed to be found near all strong \ion{Mg}{2} absorbers,
statistically. \citet{raoetal06} estimated that $\approx\!35\%$--40\%
of \ion{Mg}{2} systems with $\EWlin{2796} \ge 0.5\Ang$ and a few other
constraints (\ion{Fe}{2} $\lambda2600$ and \ion{Mg}{1} $\lambda2852$
absorption, doublet ratio) were DLAs at $z < 1.65$.

The fraction of $\EWlin{2796} \ge 1\Ang$ \ion{Mg}{2} systems
exhibiting dampled \Lya\ absorption\footnote{We estimate the number of
  absorbers by assuming we would find $\Num_{\rm ion} =
  (\dNdX{ion})\Delta X$ in a given survey path length $\Delta
  X$. Thus, we approximate the fraction of e.g., non-DLA \ion{Mg}{2}
  systems to \ion{C}{4} systems as $(\dNdX{\Mgp} -
  \dNdX{DLA})/(\dNdX{C\,IV}$).}  is consistent with the observed
$\approx\!35\%$--40\%, over the same redshift range. Overall, the
fraction grows from low-to-high redshift, becoming greater than unity
at $z \approx 4$, but the ratio of DLAs to $0.3\Ang \le \EWlin{2796} <
1\Ang$ \ion{Mg}{2} systems, though also growing, stays below 100\%.
\citet{matejekandsimcoe12} and \citetalias{zhuandmenard13} show that
\dNMgIIdX\ for these weaker systems is roughly constant from $z = 0.4
\rightarrow 5.5$. Thus, the equivalent-width limit of the DLA-tracing
\ion{Mg}{2} population may evolve with redshift.

\citet{churchilletal00a} proposed a \ion{Mg}{2} taxonomy: Classic,
Single\slash Weak, \ion{C}{4}-Deficient, Double, and DLA\slash
\ion{H}{1}-rich. They identified these classes from non-parametric
clustering analysis of 30 \ion{Mg}{2} systems with measured equivalent
widths of \Lya, \ion{Mg}{2}, \ion{Fe}{2} (or limits), and \ion{C}{4}
(or limits). Their sample came from 45 $0.4 < z < 1.4$ \ion{Mg}{2}
systems with high-resolution optical and low-resolution UV
spectroscopy and had $\EWlin{2796} < 1.8\Ang$, and $\EWlin{1548} <
2\Ang$. The Classic and Double \ion{Mg}{2} absorbers---30\% and 10\%,
respectively, of the 30---are \ion{C}{4} strong, with the latter
having the largest \EWlin{1548}\ and \EWlin{2796}, possibly because
they are two, close Classic systems. Single\slash Weak systems (23\%)
are \ion{C}{4}-weak but not remiss of absorption like the
\ion{C}{4}-Deficient group (20\%). The DLA\slash \ion{H}{1}-rich class
comprise the remaining 17\% and have the largest \EWlin{2796}\ but
\EWlin{1548}\ is comparable to the Single\slash Weak population.

The fraction of strong \ion{Mg}{2} systems, not in DLAs, relative to
\ion{C}{4} systems grows from $\approx\!10\%$ at low redshift to
roughly 40\% at $z \approx 3$, before decreasing sharply. For the
overlapping redshift range, this is consistent with the
\citet{churchilletal00a} sample, if they were to define a ``strong
\ion{Mg}{2} and \ion{C}{4} but not DLA'' class. The steep $z > 3$
decline is due to non-DLA, strong \ion{Mg}{2} systems vanishing,
possibly, as the weaker \ion{Mg}{2} systems encompass a significant
DLA population.  

However, the ratio of non-DLA \ion{Mg}{2} absorbers to non-DLA
\ion{H}{1} systems (i.e., LLSs) roughly decreases from $\approx\!50\%$
at $z = 1$ to zero at $z = 4$, though the rate of decline is steeper
at $z \gtrsim 2$ to 3. The steeper decline could be due to there being
less strong \ion{Mg}{2} systems after the $\rho_{\ast}$ peak, as
discussed previously, or because there is an increasing LLS
sub-population due to the (metal-poor) IGM at high redshift
\citep{fumagallietal13}.

We are not able to disentangle a \ion{C}{4}-Deficient, (strong)
\ion{Mg}{2} population at any redshift because \dNdX{C\,IV}\
($\EWlin{1548} \ge 0.6\Ang$) is larger than \dNMgIIdX\ ($\EWlin{2796}
\ge 1\Ang$) at all redshifts. However, in the \citet{churchilletal00a}
sample, one-half to one-third of the Single\slash Weak class have
SDSS-strength \ion{C}{4} absorption, which is roughly 10\% of their
entire sample. The shape of the \ion{C}{4} \ff{\EWr} does not evolve
significantly over $z = 1.5 \rightarrow 4.5$
\citepalias{cookseyetal13}; instead, the overall normalization changes
smoothly, driving the evolution in \dNdX{C\,IV}. Thus, the
sub-populations comprising the $\EWlin{1548} \ge 0.6\Ang$ systems must
evolve smoothly or ``conspiratorially'' to preserve the shape of
\ff{\EWlin{1548}}.


Fundamentally, the number of observed systems and their associated
ions are determined by: elemental abundances; strength and shape of the
ionizing radiation; and the spatial distribution (e.g., density,
cross-section) of the gas. The complex physics involved make
cosmological simulations, if they resolve enrichment processes in the
CGM and IGM, a powerful tool in understanding the evolution of various
absorber populations in tandem. On the flip side, results from QAL
studies are top-level constraints on ongoing and cumulative enrichment
processes and should be leveraged to assess whether cosmological
simulations reproduce reality. From the observational plane, we will
revisit the issue of the evolution of various absorber populations in
a future paper.


\section{Summary}\label{sec.summ}

We have conducted a survey for \ion{Mg}{2} systems in 79,294 SDSS DR7
quasar spectra \citep{schneideretal10}; these were chosen because they
were not BAL QSOs and had median $\langle {\rm S/N} \rangle \ge
4\,{\rm pixel}^{-1}$ in the region covering intergalactic \ion{Mg}{2}
absorption. Candidate \ion{Mg}{2} doublets were automatically
detected, and the final catalog was visually verified. This resulted
in 34,254 doublets with $\dvqso < -5000\kms$ and outside the quasar
\ion{C}{4} emission region, used for further analyses; the full
catalog and other tools for analysis (e.g., completeness grids) are
made available for the community.\footnote{See
  \url{http://igmabsorbers.info/}.}  Our Monte-Carlo completeness
tests included the effects of the automated algorithms and user bias
(e.g., the accepted false-positive rate).

Though there exist differences in methodologies, from continuum
fitting to final doublet selection, we recovered over 76\% of the
\citetalias{prochteretal06} absorbers, 80\% of
\citetalias{quideretal11}, and 72\% of \citetalias{zhuandmenard13}. We
also detect systems not in these other SDSS catalogs.

We analyzed the catalog as a whole and in 14 small redshift bins with
roughly 2500 doublets each. The equivalent-width frequency
distribution is described well by an exponential model for all
redshifts. It flattens with increasing redshift, indicating there are
more strong absorbers relative to weaker ones. Moreover, the best-fit
parameters evolve relatively smoothly from low-to-intermediate
redshift.

We compare \dNMgIIdX\ for a range of limiting equivalent widths and
find significant differential evolution with \EWlin{lim}. Namely, the
stronger \ion{Mg}{2} systems evolve more with redshift, as predicted
by the changing shape of \ff{\EWr} over time. In addition, the peak in
\dNMgIIdX\ appears to be at higher redshift for larger \EWlin{lim}. 

All SDSS studies detect the \dNMgIIdz\ increase and a peak between $z
\approx 1.5$ to 2.3. On the other hand, we do not measure a decrease
in the characteristic equivalent width, $W^{\ast}$, at $\zmgii > 1.75$
seen in \citetalias{zhuandmenard13}. However, it is difficult to
disentangle the degeneracy between $W^{\ast}$ and $\Num^{\ast}$ (i.e.,
$\alpha$ and $k$). Additionally, the \dNMgIIdX\ measurements differ by
up to 50\% due to differences in methodologies, and we discussed the
effects of Malmquist bias and offsets in rest equivalent-width
measurements. The bias arises from more absorbers scattering above
\EWlin{lim} than below, due to measurement uncertainties. A small
change in in the relative \EWlin{2796}\ ``zero point'' has a
significant impact on the absorber line density. Of course,
differences in completeness corrections can contribute greatly to
differences in survey results.

Assuming the co-moving number density of \ion{Mg}{2}-absorbing clouds
equals the co-moving number density of $B$-band-selected galaxies with
$L \ge 0.5\,L^{\ast}$, the physical cross-section of the galaxies is
estimated to grow from $\sigphys \approx 0.005\Mpc^{2}$ to
$0.015\Mpc^{2}$ over $z = 0.4 \rightarrow 2.3$. The increase in
\sigphys\ might drive the increase in \dNMgIIdX, but the evolution in
the line density might reflect a changing population of
\ion{Mg}{2}-absorbing galaxies. We also explored the effect of
cross-section scaling with the galaxy $B$-band luminosity and with
\dNMgIIdX\ for $1\Ang \le \EWr < 3\Ang$ systems, which are more likely
individual galaxy halos as opposed to intra-group gas.

A holistic approach is taken in discussing the evolution of various
absorber population, where we assume that each ion has multiple
sub-populations yielding the observed \dNdX{ion}. The intersection of
DLAs and strong \ion{Mg}{2} absorbers grows from low-to-high redshift,
but weaker ($0.3\Ang \le \EWlin{2796} < 1\Ang$) \ion{Mg}{2} systems
may better trace the DLAs at $z \gtrsim 3$.  The population of strong,
non-DLA-tracing \ion{Mg}{2} and strong \ion{C}{4} system grows from
low-to-high redshift, before decreasing sharply at $z \approx
3$. Finally the fraction of non-DLA \ion{Mg}{2} absorbers in LLSs
decreases from 50\% to zero over $z = 1 \rightarrow 4$.

This is the second paper in our Precious Metals in SDSS Quasar Spectra
project. Early versions of this catalog have already been used to
compare \ion{Mg}{2} {\it system} properties across redshift
\citep{matejeketal13} and for \ion{Mg}{2}-galaxy clustering analysis
\citep{gauthieretal13ph}. Future Precious Metals studies
include modeling \ion{Mg}{2}-absorbing cross-sections, a fairly
comparable \ion{Si}{4} catalog, and multi-ion classification analysis.


\acknowledgements The current study was funded largely by the National
Science Foundation Astronomy \& Astrophysics Postdoctoral Fellowship
(AST-1003139) and in part by MIT Undergraduate Research Opportunity
Program (UROP) Direct Funding, from the Office of Undergraduate
Advising and Academic Programming and the John Reed UROP Fund. RAS
acknowledges support from NSF grants AST-0908920 and AST-1109915, and
JXP acknowledges support from NSF AST-0709235 and AST-1010004.

We appreciate J.-R. Gauthier and H.-W. Chen's help in making our
catalog better. We thank M. Sinha for his programming help and
P. Jonsson for productive discussions regarding statistics and
programming. We gratefully acknowledge the vital role and last huzzah
(for us) of the Adam J. Burgasser Endowed Chair.

Funding for the SDSS and SDSS-II has been provided by the Alfred
P. Sloan Foundation, the Participating Institutions, the National
Science Foundation, the U.S. Department of Energy, the National
Aeronautics and Space Administration, the Japanese Monbukagakusho, the
Max Planck Society, and the Higher Education Funding Council for
England. The SDSS Web Site is \url{http://www.sdss.org/}.

The SDSS is managed by the Astrophysical Research Consortium for the
Participating Institutions. The Participating Institutions are the
American Museum of Natural History, Astrophysical Institute Potsdam,
University of Basel, University of Cambridge, Case Western Reserve
University, University of Chicago, Drexel University, Fermilab, the
Institute for Advanced Study, the Japan Participation Group, Johns
Hopkins University, the Joint Institute for Nuclear Astrophysics, the
Kavli Institute for Particle Astrophysics and Cosmology, the Korean
Scientist Group, the Chinese Academy of Sciences (LAMOST), Los Alamos
National Laboratory, the Max-Planck-Institute for Astronomy (MPIA),
the Max-Planck-Institute for Astrophysics (MPA), New Mexico State
University, Ohio State University, University of Pittsburgh,
University of Portsmouth, Princeton University, the United States
Naval Observatory, and the University of Washington.

{\it Facilities:} \facility{Sloan}

\bibliographystyle{apj} 
\bibliography{../../sdss}

\begin{thebibliography}{69}
\expandafter\ifx\csname natexlab\endcsname\relax\def\natexlab#1{#1}\fi

\bibitem[{{Bergeron}(1986)}]{bergeron86}
{Bergeron}, J. 1986, \aap, 155, L8

\bibitem[{{Bordoloi} {et~al.}(2011){Bordoloi}, {Lilly}, {Knobel}, {Bolzonella},
  {Kampczyk}, {Carollo}, {Iovino}, {Zucca}, {Contini}, {Kneib}, {Le Fevre},
  {Mainieri}, {Renzini}, {Scodeggio}, {Zamorani}, {Balestra}, {Bardelli},
  {Bongiorno}, {Caputi}, {Cucciati}, {de la Torre}, {de Ravel}, {Garilli},
  {Kova{\v c}}, {Lamareille}, {Le Borgne}, {Le Brun}, {Maier}, {Mignoli},
  {Pello}, {Peng}, {Perez Montero}, {Presotto}, {Scarlata}, {Silverman},
  {Tanaka}, {Tasca}, {Tresse}, {Vergani}, {Barnes}, {Cappi}, {Cimatti},
  {Coppa}, {Diener}, {Franzetti}, {Koekemoer}, {L{\'o}pez-Sanjuan},
  {McCracken}, {Moresco}, {Nair}, {Oesch}, {Pozzetti}, \&
  {Welikala}}]{bordoloietal11}
{Bordoloi}, R., {et~al.} 2011, \apj, 743, 10

\bibitem[{{Bouch{\'e}} {et~al.}(2012){Bouch{\'e}}, {Hohensee}, {Vargas},
  {Kacprzak}, {Martin}, {Cooke}, \& {Churchill}}]{boucheetal12}
{Bouch{\'e}}, N., {Hohensee}, W., {Vargas}, R., {Kacprzak}, G.~G., {Martin},
  C.~L., {Cooke}, J., \& {Churchill}, C.~W. 2012, \mnras, 426, 801

\bibitem[{{Bouch{\'e}} {et~al.}(2006){Bouch{\'e}}, {Murphy}, {P{\'e}roux},
  {Csabai}, \& {Wild}}]{boucheetal06}
{Bouch{\'e}}, N., {Murphy}, M.~T., {P{\'e}roux}, C., {Csabai}, I., \& {Wild},
  V. 2006, \mnras, 371, 495

\bibitem[{{Bouwens} {et~al.}(2010){Bouwens}, {Illingworth}, {Gonz{\'a}lez},
  {Labb{\'e}}, {Franx}, {Conselice}, {Blakeslee}, {van Dokkum}, {Holden},
  {Magee}, {Marchesini}, \& {Zheng}}]{bouwensetal10}
{Bouwens}, R.~J., {et~al.} 2010, \apj, 725, 1587

\bibitem[{{Bowen} \& {Chelouche}(2011)}]{bowenandchelouche11}
{Bowen}, D.~V., \& {Chelouche}, D. 2011, \apj, 727, 47

\bibitem[{{Charlton} {et~al.}(2003){Charlton}, {Ding}, {Zonak}, {Churchill},
  {Bond}, \& {Rigby}}]{charltonetal03}
{Charlton}, J.~C., {Ding}, J., {Zonak}, S.~G., {Churchill}, C.~W., {Bond},
  N.~A., \& {Rigby}, J.~R. 2003, \apj, 589, 111

\bibitem[{{Chen} {et~al.}(2010{\natexlab{a}}){Chen}, {Helsby}, {Gauthier},
  {Shectman}, {Thompson}, \& {Tinker}}]{chenetal10a}
{Chen}, H.-W., {Helsby}, J.~E., {Gauthier}, J.-R., {Shectman}, S.~A.,
  {Thompson}, I.~B., \& {Tinker}, J.~L. 2010{\natexlab{a}}, \apj, 714, 1521

\bibitem[{{Chen} {et~al.}(2010{\natexlab{b}}){Chen}, {Wild}, {Tinker},
  {Gauthier}, {Helsby}, {Shectman}, \& {Thompson}}]{chenetal10b}
{Chen}, H.-W., {Wild}, V., {Tinker}, J.~L., {Gauthier}, J.-R., {Helsby}, J.~E.,
  {Shectman}, S.~A., \& {Thompson}, I.~B. 2010{\natexlab{b}}, \apjl, 724, L176

\bibitem[{{Churchill} {et~al.}(2000){Churchill}, {Mellon}, {Charlton},
  {Jannuzi}, {Kirhakos}, {Steidel}, \& {Schneider}}]{churchilletal00a}
{Churchill}, C.~W., {Mellon}, R.~R., {Charlton}, J.~C., {Jannuzi}, B.~T.,
  {Kirhakos}, S., {Steidel}, C.~C., \& {Schneider}, D.~P. 2000, \apj, 543, 577

\bibitem[{{Churchill} {et~al.}(2013){Churchill}, {Nielsen}, {Kacprzak}, \&
  {Trujillo-Gomez}}]{churchilletal13}
{Churchill}, C.~W., {Nielsen}, N.~M., {Kacprzak}, G.~G., \& {Trujillo-Gomez},
  S. 2013, \apjl, 763, L42

\bibitem[{{Churchill} {et~al.}(1999){Churchill}, {Rigby}, {Charlton}, \&
  {Vogt}}]{churchilletal99a}
{Churchill}, C.~W., {Rigby}, J.~R., {Charlton}, J.~C., \& {Vogt}, S.~S. 1999,
  \apjs, 120, 51

\bibitem[{{Cooksey} {et~al.}(2013){Cooksey}, {Kao}, {Simcoe}, {O'Meara}, \&
  {Prochaska}}]{cookseyetal13}
{Cooksey}, K.~L., {Kao}, M.~M., {Simcoe}, R.~A., {O'Meara}, J.~M., \&
  {Prochaska}, J.~X. 2013, \apj, 763, 37

\bibitem[{{Cooksey} {et~al.}(2010){Cooksey}, {Thom}, {Prochaska}, \&
  {Chen}}]{cookseyetal10}
{Cooksey}, K.~L., {Thom}, C., {Prochaska}, J.~X., \& {Chen}, H. 2010, \apj,
  708, 868

\bibitem[{{Cool} {et~al.}(2012){Cool}, {Eisenstein}, {Kochanek}, {Brown},
  {Caldwell}, {Dey}, {Forman}, {Hickox}, {Jannuzi}, {Jones}, {Moustakas}, \&
  {Murray}}]{cooletal12}
{Cool}, R.~J., {et~al.} 2012, \apj, 748, 10

\bibitem[{{Cucchiara} {et~al.}(2013){Cucchiara}, {Prochaska}, {Zhu},
  {M{\'e}nard}, {Fynbo}, {Fox}, {Chen}, {Cooksey}, {Cenko}, {Perley}, {Bloom},
  {Berger}, {Tanvir}, {D'Elia}, {Lopez}, {Chornock}, \& {de
  Jaeger}}]{cucchiaraetal13}
{Cucchiara}, A., {et~al.} 2013, \apj, 773, 82

\bibitem[{{Fumagalli} {et~al.}(2013){Fumagalli}, {O'Meara}, {Prochaska}, \&
  {Worseck}}]{fumagallietal13}
{Fumagalli}, M., {O'Meara}, J.~M., {Prochaska}, J.~X., \& {Worseck}, G. 2013,
  \apj, 775, 78

\bibitem[{{Gabasch} {et~al.}(2004){Gabasch}, {Bender}, {Seitz}, {Hopp},
  {Saglia}, {Feulner}, {Snigula}, {Drory}, {Appenzeller}, {Heidt}, {Mehlert},
  {Noll}, {B{\"o}hm}, {J{\"a}ger}, {Ziegler}, \& {Fricke}}]{gabaschetal04}
{Gabasch}, A., {et~al.} 2004, \aap, 421, 41

\bibitem[{{Gauthier}(2013)}]{gauthier13}
{Gauthier}, J.-R. 2013, \mnras, 432, 1444

\bibitem[{{Gauthier} \& {Chen}(2011)}]{gauthieretal11}
{Gauthier}, J.-R., \& {Chen}, H.-W. 2011, \mnras, 418, 2730

\bibitem[{{Gauthier} {et~al.}(2013){Gauthier}, {Chen}, {Cooksey}, {Simcoe},
  {Seyffert}, \& {O'Meara}}]{gauthieretal13ph}
{Gauthier}, J.-R., {Chen}, H.-W., {Cooksey}, K.~L., {Simcoe}, R.~A.,
  {Seyffert}, E.~N., \& {O'Meara}, J.~M. 2013, ArXiv e-prints

\bibitem[{{Gauthier} {et~al.}(2010){Gauthier}, {Chen}, \&
  {Tinker}}]{gauthieretal10}
{Gauthier}, J.-R., {Chen}, H.-W., \& {Tinker}, J.~L. 2010, \apj, 716, 1263

\bibitem[{{Hewett} \& {Wild}(2010)}]{hewettandwild10}
{Hewett}, P.~C., \& {Wild}, V. 2010, \mnras, 405, 2302

\bibitem[{{Kacprzak} {et~al.}(2011{\natexlab{a}}){Kacprzak}, {Churchill},
  {Barton}, \& {Cooke}}]{kacprzaketal11a}
{Kacprzak}, G.~G., {Churchill}, C.~W., {Barton}, E.~J., \& {Cooke}, J.
  2011{\natexlab{a}}, \apj, 733, 105

\bibitem[{{Kacprzak} {et~al.}(2011{\natexlab{b}}){Kacprzak}, {Churchill},
  {Evans}, {Murphy}, \& {Steidel}}]{kacprzaketal11b}
{Kacprzak}, G.~G., {Churchill}, C.~W., {Evans}, J.~L., {Murphy}, M.~T., \&
  {Steidel}, C.~C. 2011{\natexlab{b}}, \mnras, 416, 3118

\bibitem[{{Kacprzak} {et~al.}(2012){Kacprzak}, {Churchill}, \&
  {Nielsen}}]{kacprzaketal12}
{Kacprzak}, G.~G., {Churchill}, C.~W., \& {Nielsen}, N.~M. 2012, \apjl, 760, L7

\bibitem[{{Komatsu} {et~al.}(2009){Komatsu}, {Dunkley}, {Nolta}, {Bennett},
  {Gold}, {Hinshaw}, {Jarosik}, {Larson}, {Limon}, {Page}, {Spergel},
  {Halpern}, {Hill}, {Kogut}, {Meyer}, {Tucker}, {Weiland}, {Wollack}, \&
  {Wright}}]{komatsuetal09}
{Komatsu}, E., {et~al.} 2009, \apjs, 180, 330

\bibitem[{{Kornei} {et~al.}(2012){Kornei}, {Shapley}, {Martin}, {Coil}, {Lotz},
  {Schiminovich}, {Bundy}, \& {Noeske}}]{korneietal12}
{Kornei}, K.~A., {Shapley}, A.~E., {Martin}, C.~L., {Coil}, A.~L., {Lotz},
  J.~M., {Schiminovich}, D., {Bundy}, K., \& {Noeske}, K.~G. 2012, \apj, 758,
  135

\bibitem[{{Lanzetta} {et~al.}(1987){Lanzetta}, {Turnshek}, \&
  {Wolfe}}]{lanzettaetal87}
{Lanzetta}, K.~M., {Turnshek}, D.~A., \& {Wolfe}, A.~M. 1987, \apj, 322, 739

\bibitem[{{Lovegrove} \& {Simcoe}(2011)}]{lovegroveandsimcoe11}
{Lovegrove}, E., \& {Simcoe}, R.~A. 2011, \apj, 740, 30

\bibitem[{{Lundgren} {et~al.}(2009){Lundgren}, {Brunner}, {York}, {Ross},
  {Quashnock}, {Myers}, {Schneider}, {Al Sayyad}, \&
  {Bahcall}}]{lundgrenetal09}
{Lundgren}, B.~F., {et~al.} 2009, \apj, 698, 819

\bibitem[{{Marchesini} {et~al.}(2007){Marchesini}, {van Dokkum}, {Quadri},
  {Rudnick}, {Franx}, {Lira}, {Wuyts}, {Gawiser}, {Christlein}, \&
  {Toft}}]{marchesinietal07}
{Marchesini}, D., {et~al.} 2007, \apj, 656, 42

\bibitem[{{Martin} \& {Bouch{\'e}}(2009)}]{martinandbouche09}
{Martin}, C.~L., \& {Bouch{\'e}}, N. 2009, \apj, 703, 1394

\bibitem[{{Matejek} \& {Simcoe}(2012)}]{matejekandsimcoe12}
{Matejek}, M.~S., \& {Simcoe}, R.~A. 2012, \apj, 761, 112

\bibitem[{{Matejek} {et~al.}(2013){Matejek}, {Simcoe}, {Cooksey}, \&
  {Seyffert}}]{matejeketal13}
{Matejek}, M.~S., {Simcoe}, R.~A., {Cooksey}, K.~L., \& {Seyffert}, E.~N. 2013,
  \apj, 764, 9

\bibitem[{{M{\'e}nard} {et~al.}(2011){M{\'e}nard}, {Wild}, {Nestor}, {Quider},
  {Zibetti}, {Rao}, \& {Turnshek}}]{menardetal11}
{M{\'e}nard}, B., {Wild}, V., {Nestor}, D., {Quider}, A., {Zibetti}, S., {Rao},
  S., \& {Turnshek}, D. 2011, \mnras, 417, 801

\bibitem[{{Misawa} {et~al.}(2008){Misawa}, {Charlton}, \&
  {Narayanan}}]{misawaetal08}
{Misawa}, T., {Charlton}, J.~C., \& {Narayanan}, A. 2008, \apj, 679, 220

\bibitem[{{Narayanan} {et~al.}(2007){Narayanan}, {Misawa}, {Charlton}, \&
  {Kim}}]{narayananetal07}
{Narayanan}, A., {Misawa}, T., {Charlton}, J.~C., \& {Kim}, T.-S. 2007, \apj,
  660, 1093

\bibitem[{{Nestor} {et~al.}(2011){Nestor}, {Johnson}, {Wild}, {M{\'e}nard},
  {Turnshek}, {Rao}, \& {Pettini}}]{nestoretal11}
{Nestor}, D.~B., {Johnson}, B.~D., {Wild}, V., {M{\'e}nard}, B., {Turnshek},
  D.~A., {Rao}, S., \& {Pettini}, M. 2011, \mnras, 412, 1559

\bibitem[{{Nestor} {et~al.}(2005){Nestor}, {Turnshek}, \& {Rao}}]{nestoretal05}
{Nestor}, D.~B., {Turnshek}, D.~A., \& {Rao}, S.~M. 2005, \apj, 628, 637

\bibitem[{{Nielsen} {et~al.}(2013){Nielsen}, {Churchill}, {Kacprzak}, \&
  {Murphy}}]{nielsenetal13}
{Nielsen}, N.~M., {Churchill}, C.~W., {Kacprzak}, G.~G., \& {Murphy}, M.~T.
  2013, \apj, 776, 114

\bibitem[{{Noterdaeme} {et~al.}(2012){Noterdaeme}, {Petitjean}, {Carithers},
  {P{\^a}ris}, {Font-Ribera}, {Bailey}, {Aubourg}, {Bizyaev}, {Ebelke},
  {Finley}, {Ge}, {Malanushenko}, {Malanushenko}, {Miralda-Escud{\'e}},
  {Myers}, {Oravetz}, {Pan}, {Pieri}, {Ross}, {Schneider}, {Simmons}, \&
  {York}}]{noterdaemeetal12}
{Noterdaeme}, P., {et~al.} 2012, \aap, 547, L1

\bibitem[{{O'Meara} {et~al.}(2013){O'Meara}, {Prochaska}, {Worseck}, {Chen}, \&
  {Madau}}]{omearaetal13}
{O'Meara}, J.~M., {Prochaska}, J.~X., {Worseck}, G., {Chen}, H.-W., \& {Madau},
  P. 2013, \apj, 765, 137

\bibitem[{{Petitjean} \& {Bergeron}(1990)}]{petitjeanandbergeron90}
{Petitjean}, P., \& {Bergeron}, J. 1990, \aap, 231, 309

\bibitem[{{Poli} {et~al.}(2003){Poli}, {Giallongo}, {Fontana}, {Menci},
  {Zamorani}, {Nonino}, {Saracco}, {Vanzella}, {Donnarumma}, {Salimbeni},
  {Cimatti}, {Cristiani}, {Daddi}, {D'Odorico}, {Mignoli}, {Pozzetti}, \&
  {Renzini}}]{polietal03}
{Poli}, F., {et~al.} 2003, \apjl, 593, L1

\bibitem[{{Prochaska} {et~al.}(2010){Prochaska}, {O'Meara}, \&
  {Worseck}}]{prochaskaetal10}
{Prochaska}, J.~X., {O'Meara}, J.~M., \& {Worseck}, G. 2010, \apj, 718, 392

\bibitem[{{Prochaska} \& {Wolfe}(2009)}]{prochaskaandwolfe09}
{Prochaska}, J.~X., \& {Wolfe}, A.~M. 2009, \apj, 696, 1543

\bibitem[{{Prochter} {et~al.}(2006{\natexlab{a}}){Prochter}, {Prochaska}, \&
  {Burles}}]{prochteretal06}
{Prochter}, G.~E., {Prochaska}, J.~X., \& {Burles}, S.~M. 2006{\natexlab{a}},
  \apj, 639, 766

\bibitem[{{Prochter} {et~al.}(2006{\natexlab{b}}){Prochter}, {Prochaska},
  {Chen}, {Bloom}, {Dessauges-Zavadsky}, {Foley}, {Lopez}, {Pettini}, {Dupree},
  \& {Guhathakurta}}]{prochteretal06grb}
{Prochter}, G.~E., {et~al.} 2006{\natexlab{b}}, \apjl, 648, L93

\bibitem[{{Quider} {et~al.}(2011){Quider}, {Nestor}, {Turnshek}, {Rao},
  {Monier}, {Weyant}, \& {Busche}}]{quideretal11}
{Quider}, A.~M., {Nestor}, D.~B., {Turnshek}, D.~A., {Rao}, S.~M., {Monier},
  E.~M., {Weyant}, A.~N., \& {Busche}, J.~R. 2011, \aj, 141, 137

\bibitem[{{Rao} \& {Turnshek}(2000)}]{raoandturnshek00}
{Rao}, S.~M., \& {Turnshek}, D.~A. 2000, \apjs, 130, 1

\bibitem[{{Rao} {et~al.}(2006){Rao}, {Turnshek}, \& {Nestor}}]{raoetal06}
{Rao}, S.~M., {Turnshek}, D.~A., \& {Nestor}, D.~B. 2006, \apj, 636, 610

\bibitem[{{Ribaudo} {et~al.}(2011){Ribaudo}, {Lehner}, \&
  {Howk}}]{ribaudoetal11a}
{Ribaudo}, J., {Lehner}, N., \& {Howk}, J.~C. 2011, \apj, 736, 42

\bibitem[{{Rigby} {et~al.}(2002){Rigby}, {Charlton}, \&
  {Churchill}}]{rigbyetal02}
{Rigby}, J.~R., {Charlton}, J.~C., \& {Churchill}, C.~W. 2002, \apj, 565, 743

\bibitem[{{Rubin} {et~al.}(2013){Rubin}, {Prochaska}, {Koo}, {Phillips},
  {Martin}, \& {Winstrom}}]{rubinetal13}
{Rubin}, K.~H.~R., {Prochaska}, J.~X., {Koo}, D.~C., {Phillips}, A.~C.,
  {Martin}, C.~L., \& {Winstrom}, L.~O. 2013, ArXiv e-prints

\bibitem[{{Rubin} {et~al.}(2010){Rubin}, {Weiner}, {Koo}, {Martin},
  {Prochaska}, {Coil}, \& {Newman}}]{rubinetal10}
{Rubin}, K.~H.~R., {Weiner}, B.~J., {Koo}, D.~C., {Martin}, C.~L., {Prochaska},
  J.~X., {Coil}, A.~L., \& {Newman}, J.~A. 2010, \apj, 719, 1503

\bibitem[{{Sargent} {et~al.}(1988){Sargent}, {Steidel}, \&
  {Boksenberg}}]{sargentetal88}
{Sargent}, W.~L.~W., {Steidel}, C.~C., \& {Boksenberg}, A. 1988, \apj, 334, 22

\bibitem[{{Schneider} {et~al.}(2005){Schneider}, {Hall}, {Richards}, {Vanden
  Berk}, {Anderson}, {Fan}, {Jester}, {Stoughton}, {Strauss}, {SubbaRao},
  {Brandt}, {Gunn}, {Yanny}, {Bahcall}, {Barentine}, {Blanton}, {Boroski},
  {Brewington}, {Brinkmann}, {Brunner}, {Csabai}, {Doi}, {Eisenstein},
  {Frieman}, {Fukugita}, {Gray}, {Harvanek}, {Heckman}, {Ivezi{\'c}}, {Kent},
  {Kleinman}, {Knapp}, {Kron}, {Krzesinski}, {Long}, {Loveday}, {Lupton},
  {Margon}, {Munn}, {Neilsen}, {Newberg}, {Newman}, {Nichol}, {Nitta}, {Pier},
  {Rockosi}, {Saxe}, {Schlegel}, {Snedden}, {Szalay}, {Thakar}, {Uomoto},
  {Voges}, \& {York}}]{schneideretal05}
{Schneider}, D.~P., {et~al.} 2005, \aj, 130, 367

\bibitem[{{Schneider} {et~al.}(2002){Schneider}, {Richards}, {Fan}, {Hall},
  {Strauss}, {Vanden Berk}, {Gunn}, {Newberg}, {Reichard}, {Stoughton},
  {Voges}, {Yanny}, {Anderson}, {Annis}, {Bahcall}, {Bauer}, {Bernardi},
  {Blanton}, {Boroski}, {Brinkmann}, {Briggs}, {Brunner}, {Burles}, {Carey},
  {Castander}, {Connolly}, {Csabai}, {Doi}, {Friedman}, {Frieman}, {Fukugita},
  {Heckman}, {Hennessy}, {Hindsley}, {Hogg}, {Ivezi{\'c}}, {Kent}, {Knapp},
  {Kunzst}, {Lamb}, {Leger}, {Long}, {Loveday}, {Lupton}, {Margon}, {Meiksin},
  {Merelli}, {Munn}, {Newcomb}, {Nichol}, {Owen}, {Pier}, {Pope}, {Rockosi},
  {Saxe}, {Schlegel}, {Siegmund}, {Smee}, {Snir}, {SubbaRao}, {Szalay},
  {Thakar}, {Uomoto}, {Waddell}, \& {York}}]{schneideretal02}
---. 2002, \aj, 123, 567

\bibitem[{{Schneider} {et~al.}(2010){Schneider}, {Richards}, {Hall}, {Strauss},
  {Anderson}, {Boroson}, {Ross}, {Shen}, {Brandt}, {Fan}, {Inada}, {Jester},
  {Knapp}, {Krawczyk}, {Thakar}, {Vanden Berk}, {Voges}, {Yanny}, {York},
  {Bahcall}, {Bizyaev}, {Blanton}, {Brewington}, {Brinkmann}, {Eisenstein},
  {Frieman}, {Fukugita}, {Gray}, {Gunn}, {Hibon}, {Ivezi{\'c}}, {Kent}, {Kron},
  {Lee}, {Lupton}, {Malanushenko}, {Malanushenko}, {Oravetz}, {Pan}, {Pier},
  {Price}, {Saxe}, {Schlegel}, {Simmons}, {Snedden}, {SubbaRao}, {Szalay}, \&
  {Weinberg}}]{schneideretal10}
---. 2010, \aj, 139, 2360

\bibitem[{{Shen} {et~al.}(2011){Shen}, {Richards}, {Strauss}, {Hall},
  {Schneider}, {Snedden}, {Bizyaev}, {Brewington}, {Malanushenko},
  {Malanushenko}, {Oravetz}, {Pan}, \& {Simmons}}]{shenetal11}
{Shen}, Y., {et~al.} 2011, \apjs, 194, 45

\bibitem[{{Simcoe} {et~al.}(2011){Simcoe}, {Cooksey}, {Matejek}, {Burgasser},
  {Bochanski}, {Lovegrove}, {Bernstein}, {Pipher}, {Forrest}, {McMurtry},
  {Fan}, \& {O'Meara}}]{simcoeetal11}
{Simcoe}, R.~A., {et~al.} 2011, \apj, 743, 21

\bibitem[{{Steidel} \& {Sargent}(1992)}]{steidelandsargent92}
{Steidel}, C.~C., \& {Sargent}, W.~L.~W. 1992, \apjs, 80, 1

\bibitem[{{Weiner} {et~al.}(2009){Weiner}, {Coil}, {Prochaska}, {Newman},
  {Cooper}, {Bundy}, {Conselice}, {Dutton}, {Faber}, {Koo}, {Lotz}, {Rieke}, \&
  {Rubin}}]{weineretal09}
{Weiner}, B.~J., {et~al.} 2009, \apj, 692, 187

\bibitem[{{Werk} {et~al.}(2013){Werk}, {Prochaska}, {Thom}, {Tumlinson},
  {Tripp}, {O'Meara}, \& {Peeples}}]{werketal13}
{Werk}, J.~K., {Prochaska}, J.~X., {Thom}, C., {Tumlinson}, J., {Tripp}, T.~M.,
  {O'Meara}, J.~M., \& {Peeples}, M.~S. 2013, \apjs, 204, 17

\bibitem[{{Willmer} {et~al.}(2006){Willmer}, {Faber}, {Koo}, {Weiner},
  {Newman}, {Coil}, {Connolly}, {Conroy}, {Cooper}, {Davis}, {Finkbeiner},
  {Gerke}, {Guhathakurta}, {Harker}, {Kaiser}, {Kassin}, {Konidaris}, {Lin},
  {Luppino}, {Madgwick}, {Noeske}, {Phillips}, \& {Yan}}]{willmeretal06}
{Willmer}, C.~N.~A., {et~al.} 2006, \apj, 647, 853

\bibitem[{{Wolfe} {et~al.}(2005){Wolfe}, {Gawiser}, \&
  {Prochaska}}]{wolfeetal05}
{Wolfe}, A.~M., {Gawiser}, E., \& {Prochaska}, J.~X. 2005, \araa, 43, 861

\bibitem[{{York} {et~al.}(2000){York}, {Adelman}, {Anderson}, {Anderson},
  {Annis}, {Bahcall}, {Bakken}, {Barkhouser}, {Bastian}, {Berman}, {Boroski},
  {Bracker}, {Briegel}, {Briggs}, {Brinkmann}, {Brunner}, {Burles}, {Carey},
  {Carr}, {Castander}, {Chen}, {Colestock}, {Connolly}, {Crocker}, {Csabai},
  {Czarapata}, {Davis}, {Doi}, {Dombeck}, {Eisenstein}, {Ellman}, {Elms},
  {Evans}, {Fan}, {Federwitz}, {Fiscelli}, {Friedman}, {Frieman}, {Fukugita},
  {Gillespie}, {Gunn}, {Gurbani}, {de Haas}, {Haldeman}, {Harris}, {Hayes},
  {Heckman}, {Hennessy}, {Hindsley}, {Holm}, {Holmgren}, {Huang}, {Hull},
  {Husby}, {Ichikawa}, {Ichikawa}, {Ivezi{\'c}}, {Kent}, {Kim}, {Kinney},
  {Klaene}, {Kleinman}, {Kleinman}, {Knapp}, {Korienek}, {Kron}, {Kunszt},
  {Lamb}, {Lee}, {Leger}, {Limmongkol}, {Lindenmeyer}, {Long}, {Loomis},
  {Loveday}, {Lucinio}, {Lupton}, {MacKinnon}, {Mannery}, {Mantsch}, {Margon},
  {McGehee}, {McKay}, {Meiksin}, {Merelli}, {Monet}, {Munn}, {Narayanan},
  {Nash}, {Neilsen}, {Neswold}, {Newberg}, {Nichol}, {Nicinski}, {Nonino},
  {Okada}, {Okamura}, {Ostriker}, {Owen}, {Pauls}, {Peoples}, {Peterson},
  {Petravick}, {Pier}, {Pope}, {Pordes}, {Prosapio}, {Rechenmacher}, {Quinn},
  {Richards}, {Richmond}, {Rivetta}, {Rockosi}, {Ruthmansdorfer}, {Sandford},
  {Schlegel}, {Schneider}, {Sekiguchi}, {Sergey}, {Shimasaku}, {Siegmund},
  {Smee}, {Smith}, {Snedden}, {Stone}, {Stoughton}, {Strauss}, {Stubbs},
  {SubbaRao}, {Szalay}, {Szapudi}, {Szokoly}, {Thakar}, {Tremonti}, {Tucker},
  {Uomoto}, {Vanden Berk}, {Vogeley}, {Waddell}, {Wang}, {Watanabe},
  {Weinberg}, {Yanny}, {Yasuda}, \& {SDSS Collaboration}}]{yorketal00}
{York}, D.~G., {et~al.} 2000, \aj, 120, 1579

\bibitem[{{Zhu} \& {M{\'e}nard}(2013)}]{zhuandmenard13}
{Zhu}, G., \& {M{\'e}nard}, B. 2013, \apj, 770, 130

\end{thebibliography}

\appendix
\section{Detailed Comparison with Previous SDSS \ion{Mg}{2}
  Catalogs}\label{appdx.prevcat}

\begin{figure*}[hbt]
  \begin{center}$
    \begin{array}{ccc}
      \includegraphics[width=0.33\textwidth]{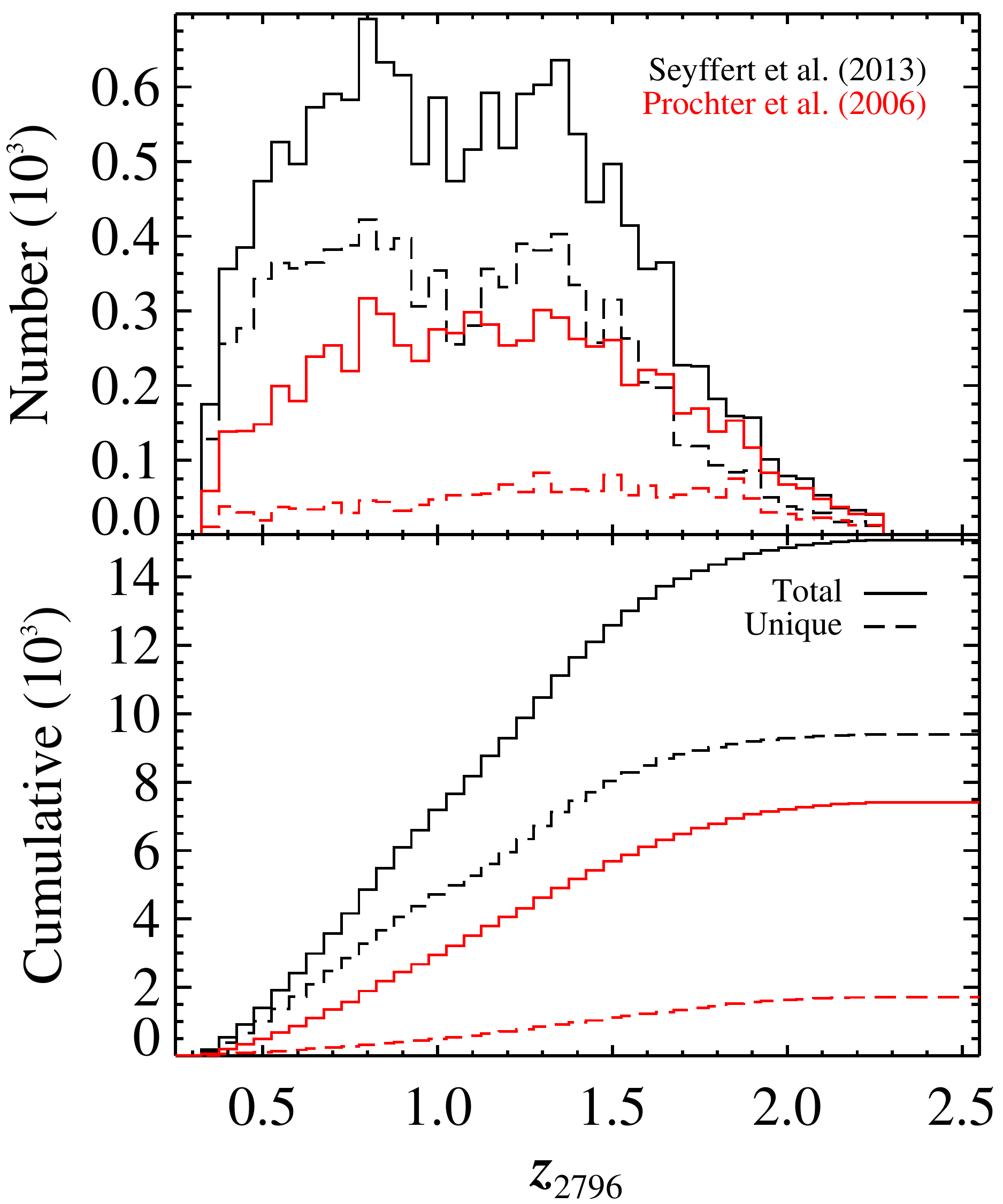} &
      \includegraphics[width=0.33\textwidth]{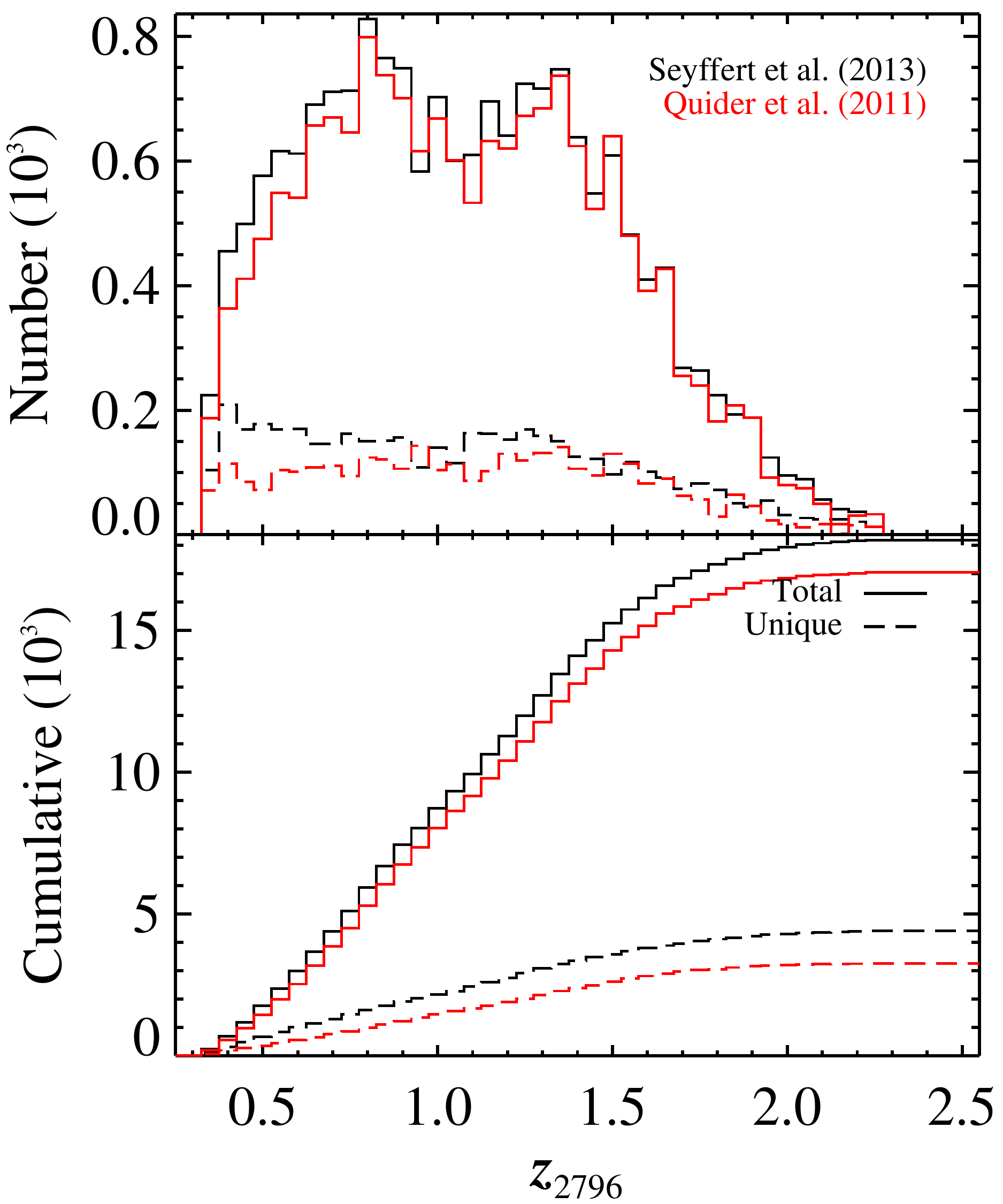}  &
      \includegraphics[width=0.33\textwidth]{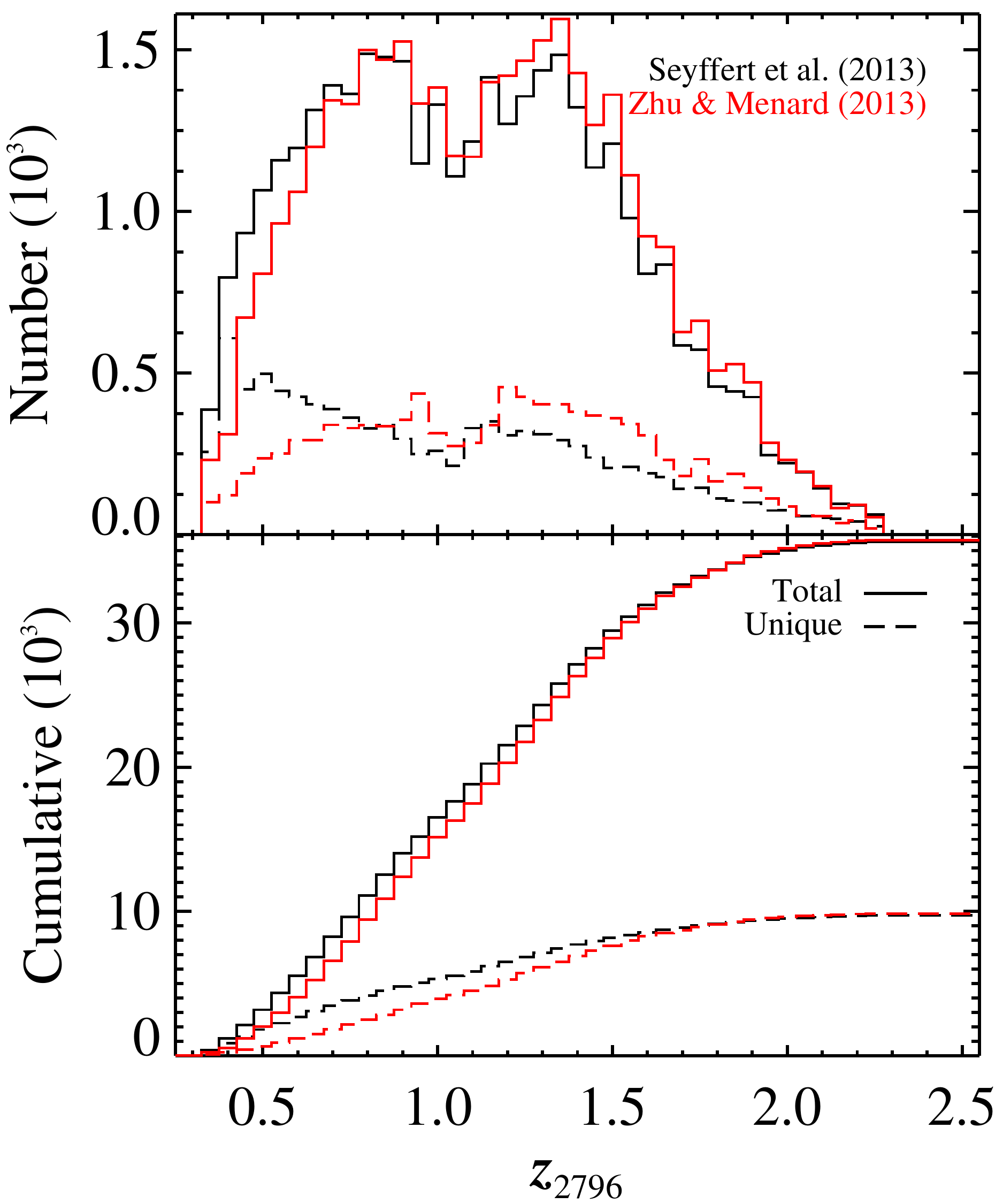}  
    \end{array}
    $\end{center}
  \caption[Comparing redshift distributions of published SDSS \ion{Mg}{2}
  catalogs.]
  {Comparing redshift distributions of published SDSS \ion{Mg}{2}
    catalogs: \citetalias{prochteretal06} (DR3, left);
    \citetalias{quideretal11} (DR4, middle); and
    \citetalias{zhuandmenard13} (DR7, right). The solid histograms
    show each catalog's full distribution (for the same data release),
    while the dashed histograms are each catalog's unique sample (our
    current catalog in black, others in red).  The hard $\EWlin{2796}
    \ge 1\Ang$ limit in \citetalias{prochteretal06} (Figure
    \ref{fig.cmpmgiiew}) drives the differences in the catalogs. For
    \citetalias{quideretal11} and \citetalias{zhuandmenard13}, the
    systems unique to the current study tend to be at lower redshift
    (black, dashed histograms).
    \label{fig.cmpmgiiz}
  }
\end{figure*}

In Section \ref{subsec.prevcat}, we summarized the differences between
four previous \ion{Mg}{2} surveys, that used the following SDSS data
releases: early \citep{nestoretal05}; third \citep{prochteretal06};
fourth \citep{quideretal11}; and seventh
\citep{zhuandmenard13}.\footnotemark[\ref{fn.lundgren}] Here we provide
specific absorber-to-absorber details.

\citetalias{nestoretal05} surveyed approximately 3700 $\zqso \ge
0.37$ sightlines in the SDSS EDR quasar catalog
\citep{schneideretal02} and visually verified 1331 \ion{Mg}{2} doublets with
$0.3\Ang \le \EWlin{2796} < 5.7\Ang$ and $0.36 < \zmgii < 2.27$. In
DR7, there remains 3809 of the 3814 total QSOs in EDR, and in those,
we detect 1425 absorbers with $\EWlin{2796} \lesssim 9\Ang$ with the same
redshift range. \citetalias{nestoretal05} did not publish their line
catalog, but it was folded in to the DR4 survey of
\citetalias{quideretal11}, which is discussed below.

\begin{figure*}[hbt]
  \begin{center}$
    \begin{array}{ccc}
      \includegraphics[trim=7cm 0cm 0cm 0cm, width=0.33\textwidth]{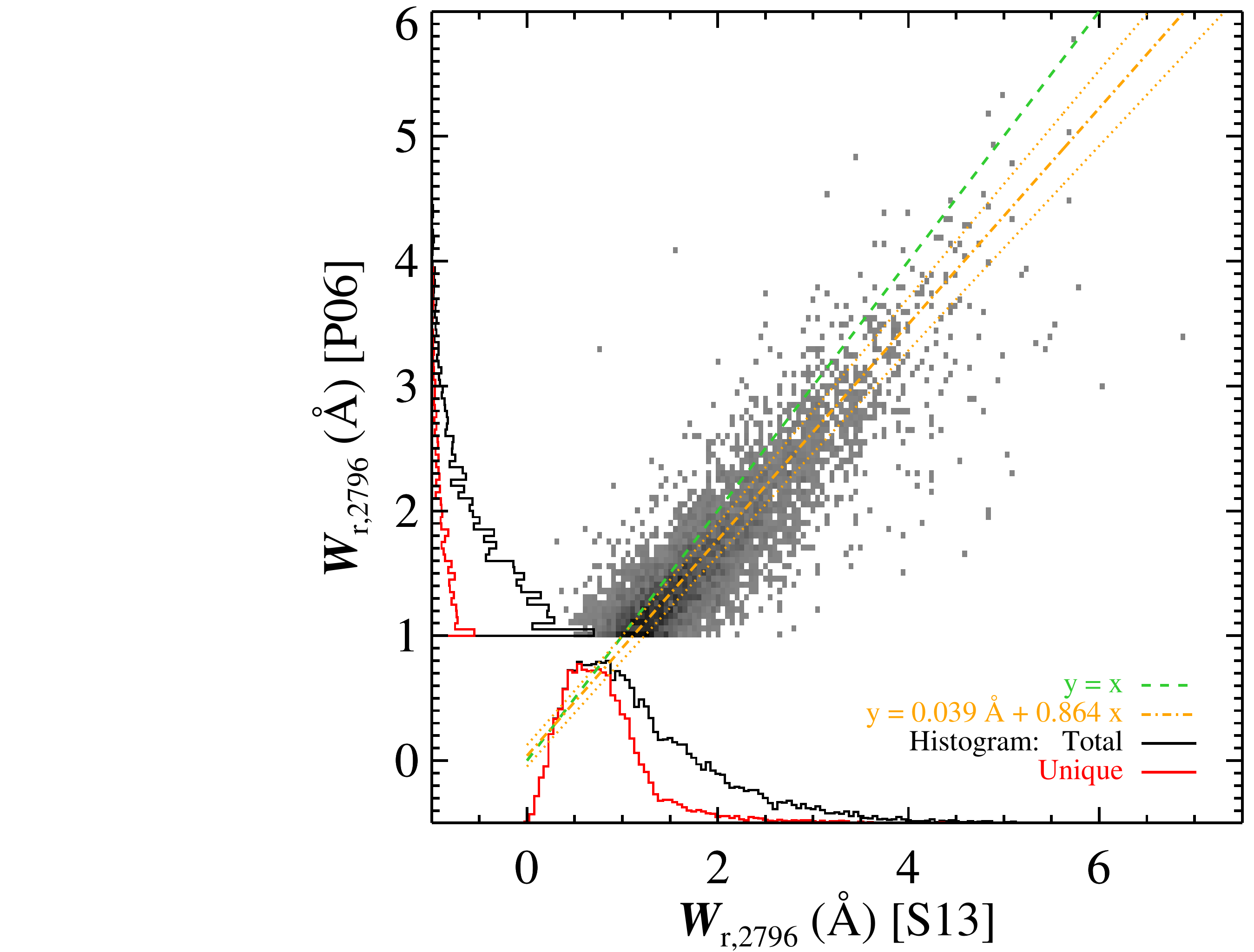}  & 
      \includegraphics[trim=7cm 0cm 0cm 0cm, width=0.33\textwidth]{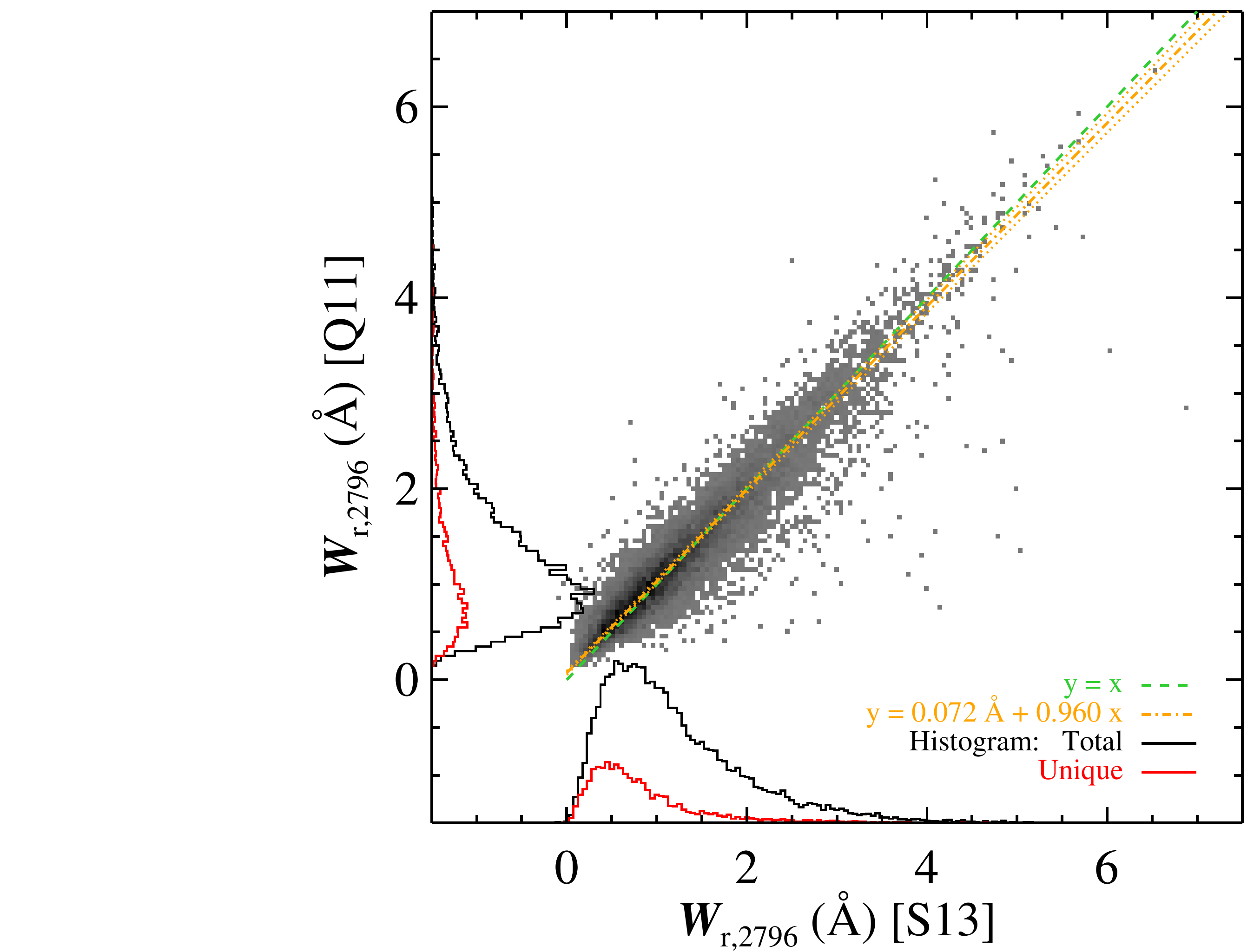}  &
      \includegraphics[trim=7cm 0cm 0cm 0cm, width=0.33\textwidth]{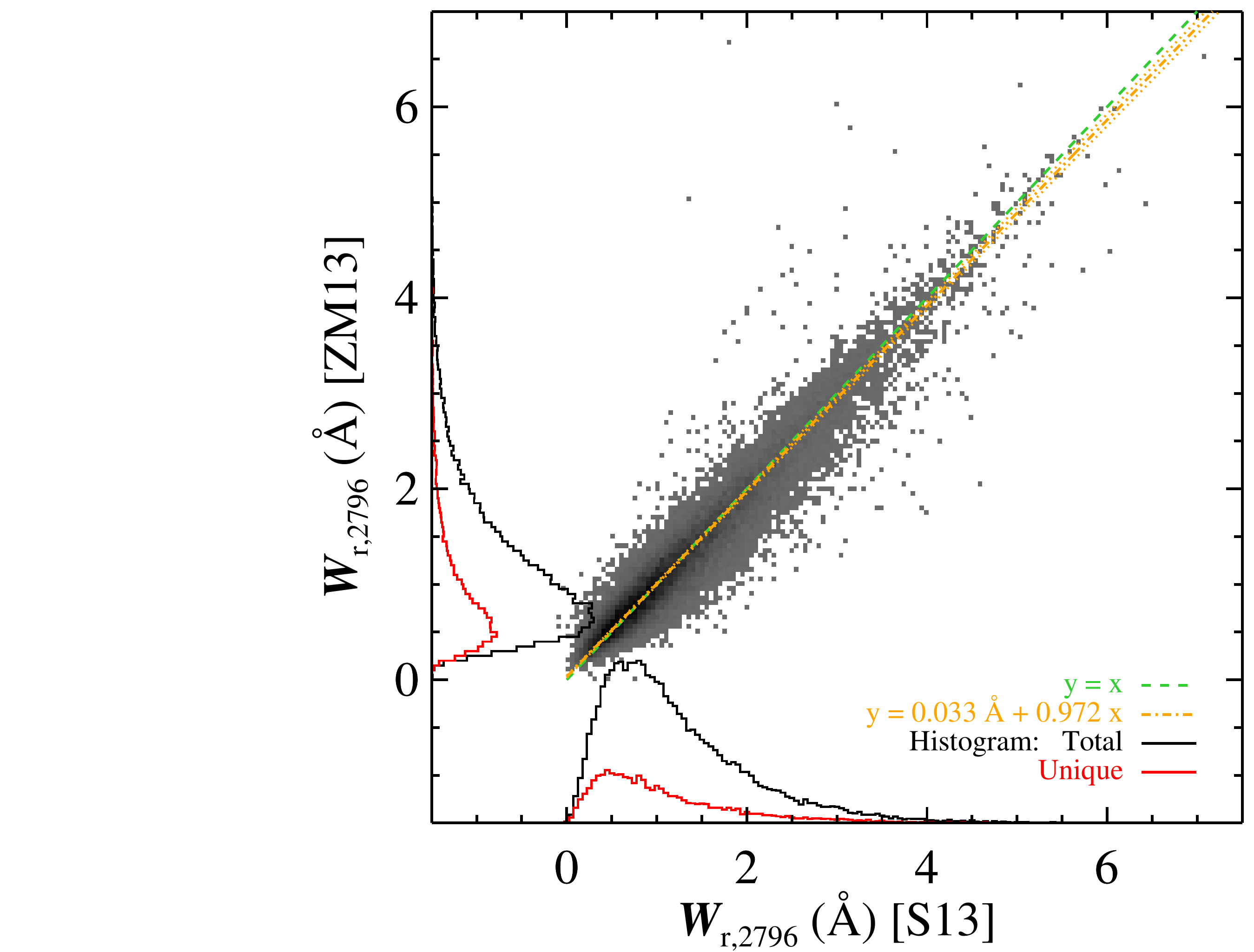}  
    \end{array}
    $\end{center}
  \caption[Comparing rest equivalent-width distributions of published SDSS
  \ion{Mg}{2} catalogs.]
  {Comparing rest equivalent-width distributions of published SDSS
    \ion{Mg}{2} catalogs. For all catalogs, the unique systems tend to
    be the weaker systems (red
    histograms). \citetalias{prochteretal06} only published
    systems with $\EWr \ge 1\Ang$ (left). They also had the largest
    discrepancy with our measured equivalent widths.  The green,
    dashed lines show the one-to-one relation. The orange, dash-dot
    lines are linear least-squares fits, including errors in both
    datasets, with three-sigma errors (orange, dotted lines). The
    black histograms show the redshift distribution of the {\it matched}
    systems from each catalog.
    \label{fig.cmpmgiiew}
  }
\end{figure*}

\citetalias{prochteretal06} surveyed 45,023 $\zqso > 0.35$ quasar
spectra in the DR3 quasar catalog \citep{schneideretal05} and
published a catalog of the resulting, visually verified 7410
doublets, with $1\Ang \le \EWlin{2796} \le 10.1\Ang$ and $0.36 < \zmgii
\le 2.28$. There are 44,545 DR3 sightlines in the DR7 catalog, and we
identified 15,083 \ion{Mg}{2} systems.
In Figures \ref{fig.cmpmgiiz} and \ref{fig.cmpmgiiew}, we compare
\zmgii\ and \EWlin{2796}\ for the intersection of the
\citetalias{prochteretal06} and our catalogs (5693 doublets) and for
the systems unique to either (1717 and 9390, respectively).  The
\citetalias{prochteretal06}-only systems are generally spread out in
redshift (Figure \ref{fig.cmpmgiiz}). \citetalias{prochteretal06}
measured equivalent widths by summing the flux absorbed in 3589\Ang\
boxes, centered at the redshift of the 2796\Ang\ line. As can be seen in
Figure \ref{fig.cmpmgiiew}, this and a systematically low continuum
fit results in an underestimation of the equivalent width for strong
systems.

We do not recover the 1717 doublets unique to
\citetalias{prochteretal06} for the following reasons (number): in
\citet{shenetal11} BAL QSO sightlines (730); $({\rm S/N})_{\rm conv} <
2.5\,{\rm resel}^{-1}$ in the 2803\Ang\ line in our normalized spectra
(588); in sightlines with $\langle {\rm S/N} \rangle < 4\,{\rm
  pixel}^{-1}$ (296); $({\rm S/N})_{\rm conv} < 3.5\,{\rm resel}^{-1}$
in the 2796\Ang\ line in our normalized spectra (65); in visual BAL
QSO sightlines (35); and within 3000\kms\ of the
\citet{schneideretal10} quasar redshift (3). Of the 9393 doublets
unique to our catalog, 6787 (72\%) have $\EWlin{2796} < 1\Ang$, which
\citet{prochteretal06} excluded by construction.

\begin{figure*}[hbt]
  \begin{center}$
    \begin{array}{cc}
      \includegraphics[trim=7cm 0cm 0cm 0cm, width=0.3\textwidth]{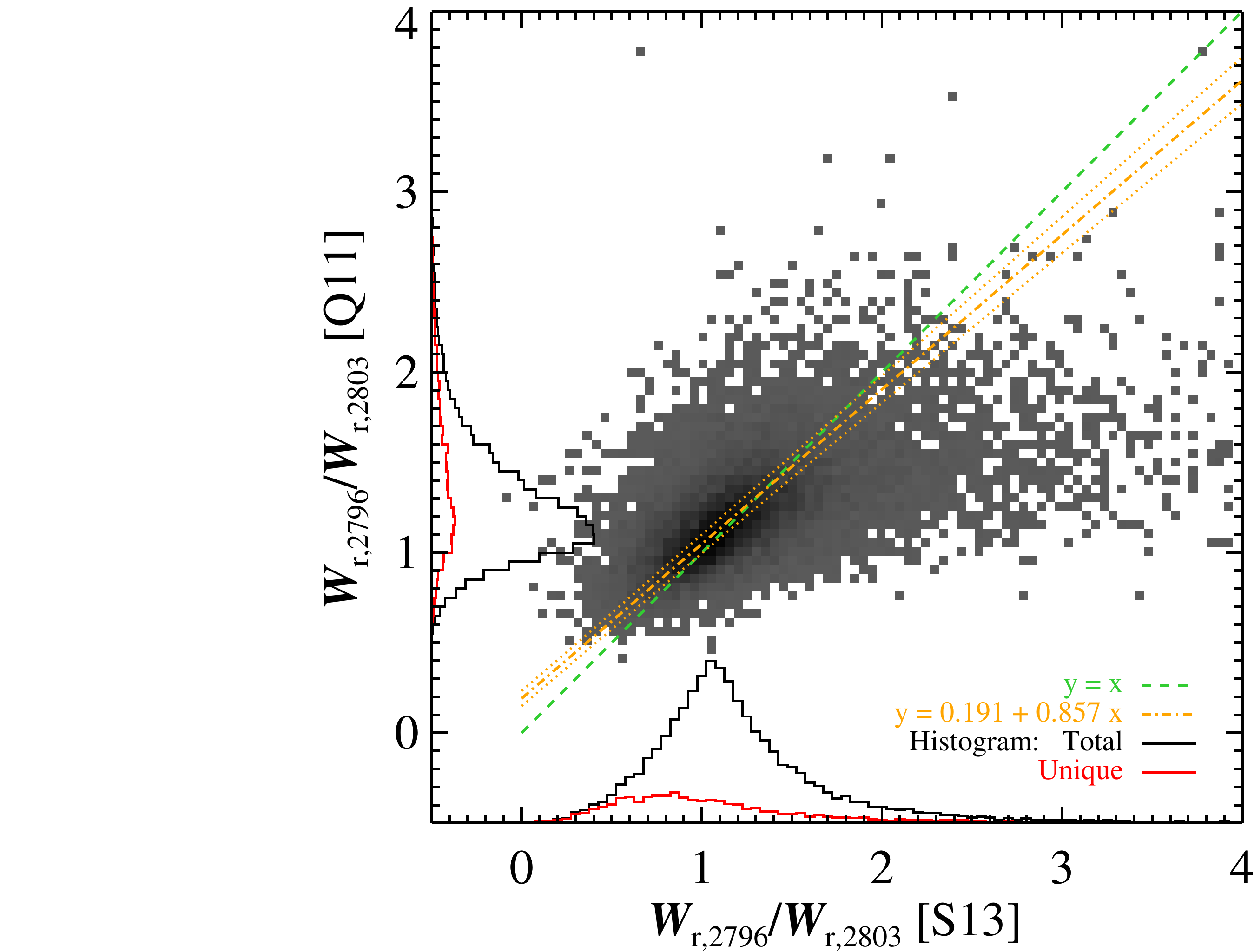}  & 
      \includegraphics[trim=7cm 0cm 0cm 0cm, width=0.3\textwidth]{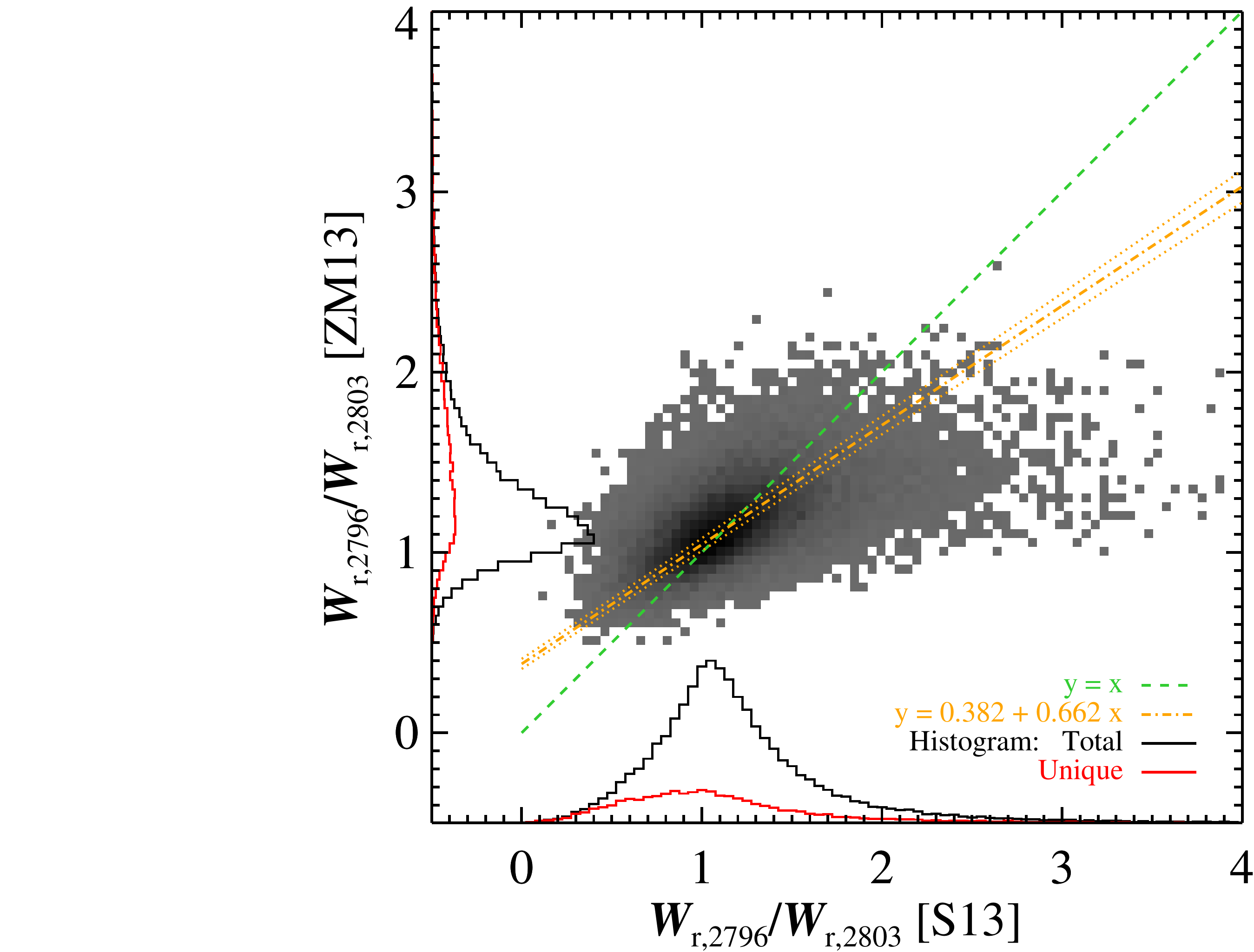}  
    \end{array}
    $\end{center}
  \caption[Comparing rest equivalent-width ratio distributions of published SDSS
  \ion{Mg}{2} catalogs.]
  {Comparing rest equivalent-width ratio distributions of published
    SDSS \ion{Mg}{2} catalogs; the colored lines and histograms are
    similar to those described in Figure
    \ref{fig.cmpmgiiew}. (\citetalias{prochteretal06} did not published
    the equivalent widths of the \ion{Mg}{2} 2803\Ang\ line.)  Though the
    equivalent widths for matched systems were largely in agreement
    (Figure \ref{fig.cmpmgiiew}), the ratios show larger
    disagreement (orange, dashed lines). The systems unique to
    \citetalias{quideretal11} and \citetalias{zhuandmenard13} tend to
    be less saturated systems (ratios $> 1$; red histograms, vertical
    axes). Since the unique systems from all catalogs were largely
    lower-$\EWlin{2796}$ systems, it appears \citetalias{quideretal11}
    and \citetalias{zhuandmenard13} were more sensitive to weak
    $\EWlin{2796}$ doublets with weaker $\EWlin{2803}$ lines, while we
    picked up on weak, saturated and\slash or blended systems, where
    blending can drive the ratio below unity.
    \label{fig.cmpmgiiewratio}
  }
\end{figure*}

\citetalias{quideretal11} surveyed 44,620 $\zqso > 0.36$ quasar
sightlines in the SDSS DR4 database, and they give the details of
their SQL query. They visually verified 17,042 \ion{Mg}{2} systems
with $0.15\Ang < \EWlin{2796} < 6.5\Ang$ and $0.36 < \zmgii <
2.29$. In DR7, there remains 44,463 of the DR4 quasars, in which we
detect 18,188
systems. 
The intersection of our catalogs yields 13,783 systems, leaving 3258
doublets unique to \citetalias{quideretal11} and 4405 to us; Figures
\ref{fig.cmpmgiiz}, \ref{fig.cmpmgiiew}, and \ref{fig.cmpmgiiewratio}
show the \zmgii, \EWlin{2796}, and \EWlin{2796}/\EWlin{2803}
distributions for matched and unique samples. There is an excess of
our unique \ion{Mg}{2} systems, at low redshift. Our \EWlin{2796}
measurements agreed well, even though \citetalias{quideretal11} used
Gaussian fits and we used boxcar summation, and systems unique to
either catalog tend to cluster at $\EWlin{2796} < 1\Ang$. There is a
significant tilt when comparing our equivalent-width ratios, and our
unique systems tend to have ratios less than unity, which mean they
have blended 2803\Ang\ lines.

We do not recover 3260 systems unique to \citetalias{quideretal11}
for the following reasons (number): in BAL QSO sightlines (1617);
$({\rm S/N})_{\rm conv} < 2.5\,{\rm resel}^{-1}$ in the 2803\Ang\ line
(1172) and $({\rm S/N})_{\rm conv} < 3.5\,{\rm resel}^{-1}$ in the
2796\Ang\ line (294) in our normalized spectra; in visual BAL QSO
sightlines (94); in sightlines with $\langle {\rm S/N} \rangle <
4\,{\rm pixel}^{-1}$ (79); and in sightlines no longer in DR7 (4).  Of
the 4403 doublets unique to our catalog, 3092 (70\%) have
$\EWlin{2796} < 1\Ang$.

\citetalias{zhuandmenard13} surveyed a total of 84,533 DR7 quasar
spectra, 569 from \citet{hewettandwild10} and the rest from
\citet{schneideretal10}. For all quasars, \citetalias{zhuandmenard13}
adopted the \citet{hewettandwild10} quasar
redshifts. \citetalias{zhuandmenard13} conducted a {\it completely
  automated} search, using \citetalias{quideretal11} as a training
set, and identified 35,752 systems with $\EWlin{2796} < 8.5\Ang$ and
$0.36 < \zmgii < 2.29$. Comparing our full catalog of 35,629
\ion{Mg}{2} systems, both catalogs agree on 25,909 systems, leaving
9843 systems unique to \citetalias{zhuandmenard13} and 9720 to
us. We show the \zmgii, \EWlin{2796}, and \EWlin{2796}/\EWlin{2803}
distributions for the matched and unique samples in Figures
\ref{fig.cmpmgiiz}, \ref{fig.cmpmgiiew}, and
\ref{fig.cmpmgiiewratio}. There is a strong trend for our unique
doublets to be at low redshift, since \citetalias{zhuandmenard13} did
not search the \ion{C}{4} ``forest,'' requiring $\Delta z > 0.02$
red-ward of the quasar \ion{C}{4} emission line. Our \EWlin{2796}
values agree well; \citetalias{zhuandmenard13} also measured the
equivalent width from Gaussian fits. The doublets unique to either
catalog peak at $\EWlin{2796} < 1\Ang$. There is a significant tilt in
the equivalent-width-ratio plane, and our unique systems are
relatively evenly distributed around a ratio of unity.

We do not recover 9843 systems unique to \citetalias{zhuandmenard13}
for the following reasons (number): $({\rm S/N})_{\rm conv} <
2.5\,{\rm resel}^{-1}$ in the 2803\Ang\ line in our normalized spectra
(5193); in BAL QSO sightlines (3075); $({\rm S/N})_{\rm conv} <
3.5\,{\rm resel}^{-1}$ in the 2796\Ang\ line in our normalized spectra
(976); in sightlines with $\langle {\rm S/N} \rangle < 4\,{\rm
  pixel}^{-1}$ (291); in visual BAL QSO sightlines (209); and in
\citet{hewettandwild10} sightlines (99).  Of the 9720 doublets unique
to our catalog, 3387 were not recovered by
\citetalias{zhuandmenard13} because they are in the \ion{C}{4} forest;
793 because they are within $\Delta z = 0.04$ of the
\citet{hewettandwild10} quasar \ion{Mg}{2} emission line; and 156
because they are coincident with Galactic \ion{Ca}{2} $\lambda\lambda
3934,3969$.

\end{document}